\documentclass[prb,aps,amssymb,shownopacs,twocolumn,superscriptaddress,10pt]{revtex4}
\usepackage{amsmath}
\usepackage{amssymb}
\usepackage{amsthm}
\usepackage{amsfonts}
\usepackage{listings}
\usepackage{enumerate}
\usepackage{latexsym}
\usepackage{psfrag}
\usepackage{bm}
\usepackage[all]{xy}
\usepackage{graphicx}
\usepackage{subfigure}
\usepackage[pdftex,colorlinks=false]{hyperref}
\usepackage{xcolor}
\usepackage{mathtools}
\usepackage{ulem}
\newcommand{\beq}{\begin{equation}}
\newcommand{\eneq}{\end{equation}}
\newcommand{\bs}[1]{\boldsymbol{#1}}

\def\be{\begin{equation}}
\def\ee{\end{equation}}
\def\ba{\begin{eqnarray}}
\def\ea{\end{eqnarray}}

\def\R{{\rm Re}}
\def\Z{\mathbb{Z}}
\def\C{\mathbb{C}}

\def\dag{\dagger}

\newcommand{\eqnref}[1]{Eq.~\eqref{#1}}
\newcommand{\punc}[1]{\,#1}

\newcommand{\secref}[1]{Sec.~\ref{#1}}

\def\mA{\mathcal{A}}
\def\mT{\mathcal{T}}
\def\mU{\mathcal{U}}

\def\beq{\begin{equation}}
\def\eeq{\end{equation}}
\def\barray{\begin{eqnarray}}
\def\earray{\end{eqnarray}}

\font\upright=cmu10 scaled\magstep1
\def\stroke{\vrule height8pt width0.4pt depth-0.1pt}

\def\Zmath{\mathbb{Z}}
\def\Qmath{\vcenter{\hbox{\upright\rlap{\rlap{Q}\kern
                   3.8pt\stroke}\phantom{Q}}}}
\def\Nmath{\vcenter{\hbox{\upright\rlap{I}\kern 1.7pt N}}}
\def\Cmath{\vcenter{\hbox{\upright\rlap{\rlap{C}\kern
                   3.8pt\stroke}\phantom{C}}}}
\def\Rmath{\vcenter{\hbox{\upright\rlap{I}\kern 1.7pt R}}}
\def\Z{\ifmmode\Zmath\else$\Zmath$\fi}
\def\Q{\ifmmode\Qmath\else$\Qmath$\fi}
\def\N{\ifmmode\Nmath\else$\Nmath$\fi}
\def\C{\ifmmode\Cmath\else$\Cmath$\fi}
\def\R{\ifmmode\Rmath\else$\Rmath$\fi}

\input{epsf}

\begin{document}

\tolerance 10000

\newcommand{\cbl}[1]{\color{blue} #1 \color{black}}

\newcommand{\vk}{{\bf k}}

\title{Boson Condensation in Topologically Ordered Quantum Liquids}

\author{
Titus Neupert}
\address{
 Princeton Center for Theoretical Science, Princeton University, Princeton, New Jersey 08544, USA
}

\author{
Huan He}
\address{
Department of Physics, Princeton University, Princeton, New Jersey 08544, USA
}

\author{
Curt von Keyserlingk}
\address{
 Princeton Center for Theoretical Science, Princeton University, Princeton, New Jersey 08544, USA
}

\author{
Germ\'an Sierra}
\address{
Instituto de F\'isica Te\'orica, UAM-CSIC, Madrid, Spain
}

\author{
B. Andrei Bernevig}
\address{
Department of Physics, Princeton University, Princeton, New Jersey 08544, USA
}

\begin{abstract}
Boson condensation in topological quantum field theories (TQFT) has been previously investigated through the formalism of Frobenius algebras and the use of vertex lifting coefficients. While general, this formalism is physically opaque and computationally arduous: analyses of TQFT condensation are practically performed on a case by case basis and for very simple theories only, mostly not using the Frobenius algebra formalism. In this paper we provide a new way of treating boson condensation that is computationally efficient. With a minimal set of physical assumptions, such as commutativity of lifting and the definition of confined particles, we can prove a number of theorems linking Boson condensation in TQFT with chiral algebra extensions, and with the factorization of completely positive matrices over $\mathbb{Z}_+$. We present numerically efficient ways of obtaining a condensed theory fusion algebra and $S$ matrices; and we then use our formalism to prove several theorems for the $S$ and $T$  matrices of simple current condensation and of theories which upon condensation result in a low number of confined particles. We also show that our formalism easily reproduces results existent in the mathematical literature such as the noncondensability of 5 and 10 layers of the Fibonacci TQFT.
\end{abstract}

\date{\today}

\maketitle

\section{Introduction}

In two spatial dimensions the Pauli principle generalizes, allowing for anyonic particles with quantum statistics different from that of fermions and bosons\cite{W82}. While such anyons do not appear as free particles in nature, they can occur as emergent excitations in quasi two-dimensional fractional quantum Hall systems\cite{FQH1,FQH2}, and theoretically in other states of quantum matter, such as the toric code and its generalizations\cite{Kitaev,Kitaev06}. A state with anyonic excitations is called topologically ordered\cite{WWZ89,Wen89,Wen90,Wen15,CGW10,GW09,Wen04} and represents a new paradigm in condensed matter physics with far reaching potential technological applications in quantum computation\cite{TQC}. 

Prior to the discovery of topological order, it was well known that bosons can macroscopically occupy a single quantum state, a fact which allows for the possibility of a Bose-Einstein condensation phase transition. In a topologically ordered phase, bosons are more complicated particles: They can have nontrivial braiding behavior with other anyons\cite{Kitaev06,Bonderson07}, and even more exotically they can carry nonlocal internal degrees of freedom\cite{Kitaev06}, in which case they are called non-Abelian bosons. Notwithstanding, such bosons can sometimes condense\cite{Bais1,Bais2,Bais3,YJW12,BW10,BSS12,BSS11,HW13,HW15b,Bais06,Bais07,Bais02}. It is then natural to ask how this condensation affects the topological order, namely, what is the fate of the other anyons in the phase. The answer is that anyon condensation induces transitions between different topologically ordered phases in such a way that universal properties of the anyons of the condensed phase can be inferred from those of the initial phase, together with a list of condensed bosons. This framework of anyon condensation transitions found many applications in the study of topological order,~\cite{BDP92,MLSA13,GHW14,SDMSV14,SDSV13,BMD08} in particular in solving the question of bulk-boundary correspondence~\cite{BK98,BSW10,BSW11,KS11,L13,K14}, or in deducing the universal properties of domain walls~\cite{BSH09,LWW15,Kawahigashi,HW15a,BH15,WW15,KK12,GS09,BS10} and other external defects~\cite{Bombin10,BQ12,BJQ13a,BJQ13b,BJK13c,YW12,TRC14,TLF15}.

The universal aspects of topologically ordered phases are captured by topological quantum field theories\cite{WittenJonesPoly}. Among these, the axiomatic approach of category theory~\cite{FFRC06,LW05,LW14,KW14,KWZ15}, more concretely the formulation of modular tensor categories (MTCs), is particularly powerful and, to our knowledge, provides a complete characterization of topological order in two-dimensional space.~\cite{Kitaev06,Bonderson07} At a basic level, MTC's are characterized by the types of anyons that appear in the phase as well as their interrelations in the form of fusion and braiding information, the so-called ``$F$ moves'' and ``$R$ moves''.

In correspondence with the different descriptions of topological order itself, several formulations of anyon condensation were developed. In the context of MTCs, the phase after condensation is found by studying commutative separable Frobenius algebras~\cite{FRS01,FS03,FRS03,FSS11,M03} of the initial theory.~\cite{KO01,L14} Bais and Slingerland translated this procedure into the language of anyon models~\cite{Bais1}, but their formulation did not give a systematic method for determining properties of the phase after condensation. This was later achieved by Eli\"{e}ns et al.\ in Ref.~\onlinecite{Bais2} via a diagrammatic formulation of the condensation problem that makes use of the so-called vertex lifting coefficients. These allowed them to embed the fusion and braiding processes of the condensed phase in the initial anyon model. However, all these approaches fall short of providing an algorithmic formulation of boson condensation in a way that could, for example, be implemented in a computer algebra program allowing for systematic studies of possible condensations.

In this paper, we reformulate the problem of boson condensation in anyon models axiomatically and purely algebraically. The resulting formalism is based on a small number of natural assumptions such as the commutativity of fusion and condensation as well as an assumption about the topological spins of the anyons after condensation. Our approach puts the modular matrices $S$ and $T$ of the initial anyon model center stage, instead of focusing on the $F$ and $R$ moves, which are the key objects of interest for the diagrammatic approach.~\cite{Bais2} The $F$ and $R$ moves are in general notoriously hard to compute even for relatively simple theories. Our goal is to find the modular matrices $\tilde{S}$ and $\tilde{T}$ of the final theory after condensation. Using our algebraic formulation, we propose an algorithm that determines all possible condensation instabilities of an anyon model and can be efficiently implemented on a computer. We solve for the condensation via a series of linear algebra problems, involving the factorization of nonnegative integer matrices.

Besides its utility for computer-aided calculations, our algebraic formulation of condensation also facilitates analytical derivations. As an example, we discuss layer constructions of topologically ordered states and easily reproduce the known result that 5 and 10 layers of the Fibonacci anyon model cannot undergo a condensation transition. 

This paper is structured as follows. In Sec.~\ref{sec.def} we formulate the condensation problem along with the axioms relating to fusion rules. In the following Sec.~\ref{sec.Mmatrix}, we present the assumptions that allow us to deduce the braiding properties of the theory after the condensation transition, and several implications are derived. In Sec.~\ref{sec.MTC}, we derive central equations which constrain $\tilde{S}$ and $\tilde{T}$. Subsequently, we show in Secs.~\ref{sec.simplecurrent} and~\ref{sec: One confined particle theories} that a weaker set of axioms suffices, if the condensate consists of so-called simple currents and if only one particle is confined through the condensation transition, respectively. We formulate an algorithm for solving the condensation problem in Sec.~\ref{sec.examples}. The final Sec.~\ref{sec: layer construction} gives examples of condensation transitions in multi-layered anyon models and discusses obstructions to boson condensation in 5 and 10 layers with Fibonacci anyons.  We have included eight  appendices containing 
brief summaries of MTCs (Appendix~\ref{app.MTC})  and chiral algebras (Appendix~\ref{sec.chiralextension}),  mathematical proofs of the results explained in the main text
(Appendices~\ref{app.QutmDim},  \ref{app.lemma}, \ref{app.beta}) and further examples of condensation (Appendices~\ref{app: SU2 condensation}, \ref{sec: 4 layers Ising}, \ref{app: condensation of D_2}).

\section{Definitions and Assumptions}
\label{sec.def}

In this section we present the formalism underpinning anyon condensation, following Refs.~\onlinecite{Bais1} and~\onlinecite{Bais2} closely. Our discussion is self-contained with respect to the previous literature on anyon condensation, but assumes that the reader is familiar with the basic concepts of MTCs \cite{Kitaev} (see Appendix~\ref{app.MTC} for a short review).

The input for our approach to anyon condensation is a MTC $\mA$ (the uncondensed theory), and a set of restriction and lifting coefficients, which relate the particle excitations in $\mA$ to those in $\mT$ (the condensed theory). In general, $\mT$ is only a fusion category, because it may contain some excitations which are confined by the surrounding condensate. Projecting out these confined excitations, we are left with a deconfined condensed MTC that we denote as $\mU$. Our goal is to find possible MTCs $\mU$ given $\mA$ and some basic information about the condensate, such as which bosons condense.

In what follows, we will consider the Bose condensation of a collection of bosons in the original theory $\mA$. This collection of anyons is called the condensate and becomes part of the vacuum in the new intermediate fusion category $\mT$. In condensing these bosons, a generic anyon $a\in \mA$ will become (or ``restrict to'') a superposition of particles $t \in \mT$
\begin{eqnarray} \label{eq:definerestriction}
a \xmapsto{r} \sum_{t\in \mT} n_a^t t,\quad\forall a\in\mA
\end{eqnarray}
with the coefficients $n_{a}^{t}\in \mathbb{Z}_{\geq 0}$, where we assume that $n^t_a=n^{\bar{t}}_{\bar{a}}$ and bars denote antiparticles (see Appendix~\ref{app.MTC}). Equation~\eqref{eq:definerestriction} defines the ``restriction map''. We will also use the phrase ``$a$ restricts to $\sum_{t} n_{a}^{t} t$'' to describe \eqnref{eq:definerestriction}. It is possible that more than one particle $t$ appears on the righthand side of \eqnref{eq:definerestriction}, in which case we say that ``$a$ splits into $\sum_{t} n_{a}^{t} t$''. Condensed particles (bosons $b$ in the condensate) have the additional special property that $n^{\varphi}_{b}\neq 0$, where $\varphi$ is the vacuum particle in $\mT$, that is, their restriction contains the identity of the new $\mT$ theory. If $n_b^\varphi\neq 0$, then $n_b^\varphi=n_{\bar{b}}^\varphi$, i.e., both the boson and its antiparticle must condense at the same time.

The reverse (or, more precisely, adjoint) operation to restriction is called ``lifting''. For a particle $t\in\mT$, all the particles in $\mA$ which restrict to $t$ are defined to be the lifts of $t$. The lifting coefficients are the same $n_a^t$ that we used in defining the restriction. Formally, lifting is defined by
\begin{eqnarray}
t \xmapsto{l} \sum_{a\in\mA} n_a^t a,\quad\forall t\in\mA.
\end{eqnarray}
Finally, we define particles in $\mT$ whose lifts do not share a common topological spin  $\theta_a$  as confined, that is
\begin{eqnarray}
t:  {\rm confined} \; \Leftrightarrow  \; \exists  \, a, b \;  {\rm such} \, {\rm that} \;   n^t_a n^t_b \neq 0  \; {\rm with} \;   \theta_a \neq \theta_b  . 
\label{confined}
\end{eqnarray}
Conversely, the deconfined particles in $\mT$ are the particles whose liftings do share a common topological spin, which becomes identified with the spin of the deconfined particle, that is 
\begin{eqnarray}
t:  {\rm deconfined} \; \Leftrightarrow  \; \forall   \, a,  b \;  {\rm such} \, {\rm that} \;   n^t_a n^t_b \neq 0  \; {\rm then} \;   \theta_a =  \theta_b  . 
\label{deconfined}
\end{eqnarray}
Obviously, any  particle $t \in  \mT$  is either deconfined ($t \in {\cal U}$) or confined  ($t \in {\cal T/U}$). 
With these definitions in place, we now make a fundamental assumption from which we will derive the structure of the theory after condensation. 
We assume that the restriction $\mA\rightarrow \mT$ commutes with fusion. 
This is represented by the diagram
\begin{displaymath}\label{diag.commuteRstrFus}
    \xymatrix{
        \mA\otimes\mA \ar[r]^{f} \ar[d]_{r \otimes r} & \mA \ar[d]^{r} \\
        \mT\otimes\mT \ar[r]_{f}       & \mT }
\end{displaymath}
in which $f$ represents fusion and $r$ represents restriction. 

More explicitly, the commuting diagram can be written in terms of anyon basis 
\begin{eqnarray}\label{eq.commuteRstrFus}
\sum_{r,s\in\mT} n^r_a n^s_b \tilde{N}^t_{rs}=\sum_{c\in\mA} N^c_{ab} n^t_c,
\end{eqnarray}
where $N^c_{ab}$ and $\tilde{N}^t_{rs}$ are the fusion coefficients in $\mA$ and $\mT$, respectively. 
This elementary constraint is surprisingly restrictive. For instance, it immediately leads us to two constraints on the quantum dimensions of particles in the $\mA$ and $\mT$ theories (see Appendix \ref{app.QutmDim})
\begin{subequations}
\label{eq.qtdim}
\begin{eqnarray}
\label{eq.qtdimAT}
d_a&=&\sum_{r\in\mT} n_a^r d_r,\quad\forall~a\in\mA,\\
\label{eq.qtdimTA}
d_t&=&\frac{1}{q}\sum_{a\in\mA} n^t_a d_a,\quad\forall~t\in\mT,
\end{eqnarray}
\end{subequations}
where $q:=\sum_a n^\varphi_a d_a$. Diagrammatically, \eqnref{eq.qtdimTA} is
\begin{equation}
\vcenter{\hbox{\includegraphics[scale=0.29]{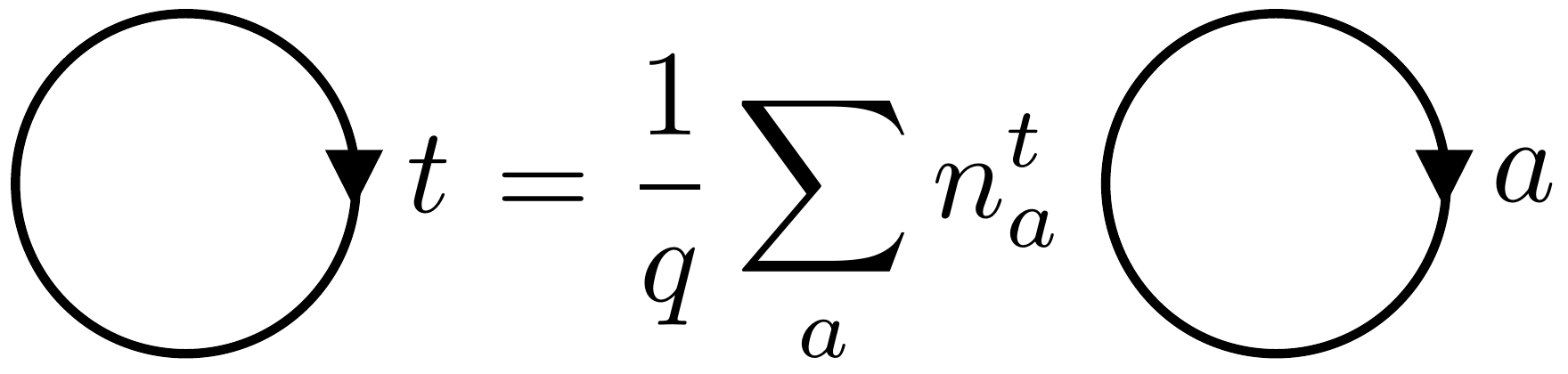}}}
\end{equation}

It will also be useful to define the quantity
\begin{equation}
\beta_t:= \sum_{a\in\mA} \theta_a d_a n^t_a,
\end{equation}
where $\theta_a$ is the topological spin of $a\in\mA$.
Given a particle $t\in\mU$, 
it follows from the aforementioned definition of a deconfined particle Eq.(\ref{deconfined})  
\begin{equation}\label{a1} 
\beta_{t} =  q d_{t}\theta_t, \quad \forall t\in \mU,
\end{equation}
as a useful corollary to \eqnref{eq.qtdimTA}.

\section{The Condensation Matrix $M_{ab}$}
\label{sec.Mmatrix}
 
So far, our formalism does not differ appreciably from that of Refs.~\onlinecite{Bais1} and~\onlinecite{Bais2}. However, in what follows we opt to not introduce the so-called ``vertex lifting coefficients'' on which the approach of \onlinecite{Bais2} is based. Instead, we find that we can extract a surprising amount of information from supplementing the algebraic relations in \secref{sec.def} with two additional assumptions. First, by assumption, we are only interested in cases where $\mU$ is a TQFT, so that its anyons form a braided fusion category. Second, we assume that
\begin{equation}
\beta_{t}=0,\quad\forall t \in \mT/\mU,
 \label{a2} 
\end{equation}
where $t \in\mT/\mU$ runs over all confined anyons. To motivate this equation let us pictorially  represent the lefthand side of Eq.~\eqref{a2} as
\begin{equation}
\vcenter{\hbox{\includegraphics[scale=0.29]{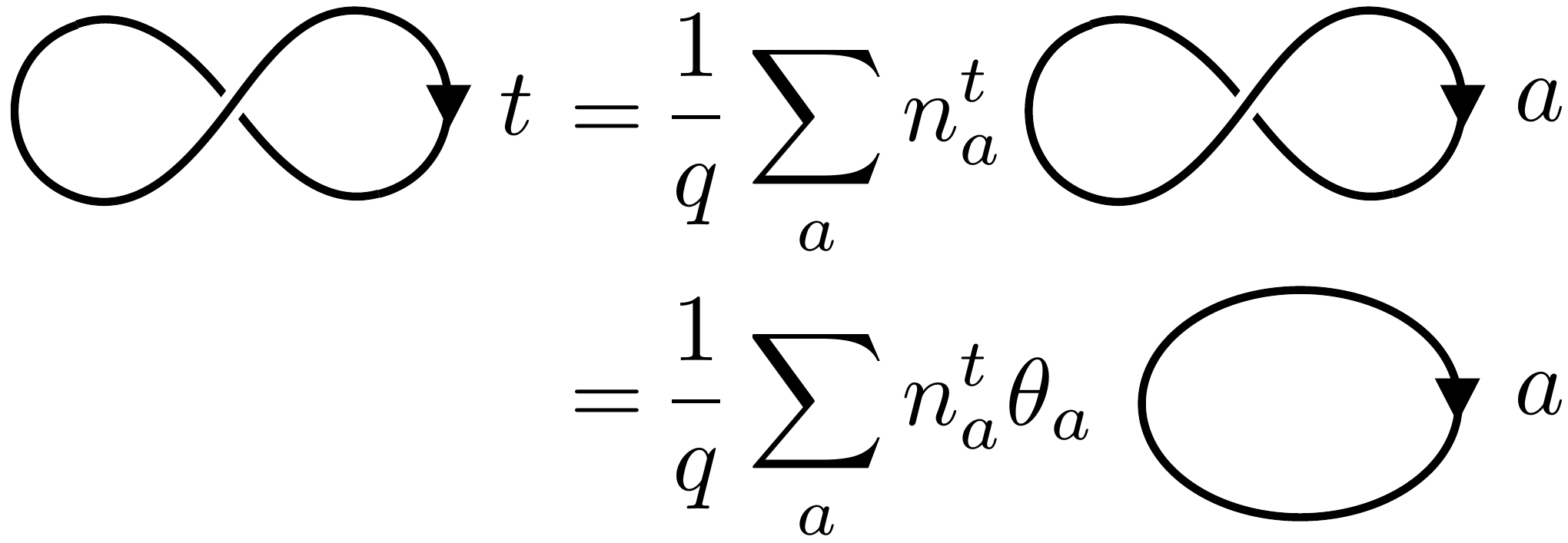}}}
\end{equation}
where a particle $t$  is braided around itself. This process is equivalent to braiding 
the lifts $a$ of the particle $t$ (namely $n^t_a \neq 0$). Each of these braidings 
is given by the phase $\theta_a$, while the loop with particle $a$ is equal to the
quantum dimension $d_a$. The result we obtain is the quantity $\beta_t$ which
we assume vanishes when $t \in {\cal T}/{\cal U}$ as confined particles cannot form a braided category. While in general we assume \eqnref{a2} in addition to the statements in \secref{sec.def}, we note that it can be derived from the assumptions underpinning the approaches in Kirillov-Ostrik~\cite{KO01}, Kong~\cite{L14} and Eli\"{e}ns et al.~\cite{Bais2} In certain special cases we can show that \eqnref{a2} follows from the assumptions: in \secref{sec.def}, e.g., we do so for the so-called simple current condensates (see \secref{sec.simplecurrent}).

\begin{subequations}
With these additional assumptions in place, we define some useful quantities. The vacuum component $t=\varphi$ of \eqnref{eq.commuteRstrFus} will be a central object in our analysis, the left hand side of which reads: 
\begin{eqnarray}\label{eq: M equation for T}
&&M'_{ac}:=\sum_{t \in\mT} n_a^t n_c^t \ \ \left( = \sum_{b\in\mA} N_{ab}^c n_b^\varphi \right)  \punc{,}
\end{eqnarray}
as will be
\begin{eqnarray}\label{eq: M equation for U}
M_{ac}:=\sum_{t\in\mU} n_a^t n_c^t \punc{.}
\end{eqnarray}
Notice how the two above definitions of the matrices $M'$ and $M$ with nonnegative integer entries differ subtly but crucially: The expression for $M'$ involves a summation over $\mT$ while that for $M$ involves a summation over the deconfined condensed theory $\mU$. 

The matrices, which can be factorized as in Eq.~\eqref{eq: M equation for T} and \eqref{eq: M equation for U}, are called completely positive matrices over the ring of positive integers. We will discuss completely positive matrix factorization later in the paper. In the following sections, we will demonstrate two important properties of $M$, namely $[M,S]=[M,T]=0$, where $S$ and $T$ are modular matrices of the $\mA$ theory. For a discussion of the role of the matrix $M$ in CFTs, we refer the reader  to Appendix~\ref{sec.chiralextension}.
\end{subequations}

\subsection{Proof that $M$ commutes with $T$ matrix of the $\mA$ theory} 

In the following, we will prove that the $M$ matrix we defined in Eq.~\eqref{eq: M equation for U} commutes with the modular $T$ matrix of the $\mA$ theory. The $T$ matrix of the $\mA$ theory is $T_{ab} = \theta_a \delta_{ab}$. Note that
\begin{equation}
\begin{split}
[M, T]_{ac}=& 
\sum_{b \in \mA} \left(M_{ab} T_{bc} - T_{ab} M_{bc}\right)	\\
= & \; M_{ac} \theta_c - M_{ac} \theta_a		\\
=& \sum_{t \in \mU} n_a^t n_c^t(\theta_c- \theta_a).
\end{split}
\label{eq: proof M commutes with T}
\end{equation}
Since $t\in \mU$, the spins of all the lifts of $t$ in $\mA$ are the same, hence $\theta_a = \theta_c$ and each term in the final line vanishes identically. It follows that
\begin{equation}\label{eq:[M,T]=0}
[M,T]=0\punc{.}
\end{equation}
Note that this is not valid if the sum in the last line of \eqnref{eq: proof M commutes with T} was not restricted to the $\mU$ theory, i.e., $[M', T]\ne 0$, with $M'$ defined in Eq.~\eqref{eq: M equation for T}. 

\subsection{Proof that $M$ commutes with $S$ matrix of the $\mA$ theory}

In this section, we will prove that the $M$ matrix commutes with the modular $S$ matrix of $\mA$ theory 
\begin{equation}\label{eq:[M,S]=0}
[M,S]=0\punc{.}
\end{equation}

We start from the expression of the $S$ matrix for a braided fusion category $\mA$ (e.g., see Ref.~\onlinecite{BernevigNeupert15}) 
\begin{equation}
S_{cb} = \frac{1}{D_\mA} \sum_{x\in\mA} N_{c \bar{b}}^x \frac{\theta_x}{\theta_c \theta_b} d_x,
\label{eq: def S}
\end{equation} 
where $D_\mA$ is the total quantum dimension of the $\mA$ theory and $\bar{b}$ denotes the antiparticle of $b$. From the definition of $M$, we express the commutator $[M,S]$ as
\begin{widetext}
\begin{equation}
\begin{split}
[M, S]_{ab}&= 
 \frac{1}{D_\mA \theta_a \theta_b}   \sum_{c,x \in\mA}  \sum_{t\in\mU} \theta_x d_x \left(n_a^t n_c^t   N_{c \bar{b}}^x   - n_b^t n_c^t N_{a\bar{c}}^x \right)	
\\
&= \frac{1}{D_\mA \theta_a \theta_b} \sum_{x\in\mA}  \sum_{t \in \mU} \theta_x d_x \left[n_a^t  \left(\sum_{c \in \mA} n_c^t   N_{c \bar{b}}^x\right) - n_b^t \left(\sum_{c\in\mA} n_c^t N_{a\bar{c}}^x \right)\right]
\\
&=\frac{1}{D_\mA \theta_a \theta_b} \sum_{x\in\mA}  \sum_{t\in\mU} \theta_x d_x \left[n_a^t  \left(\sum_{c\in\mA} n_c^t   N_{bx}^c\right)   - n_b^t \left(\sum_{c\in\mA} n_c^t N_{a\bar{x}}^c \right) \right]
.
\end{split}
\end{equation} 
In the first line we  have used that if $n^t_a n^t_c \neq 0$ with $t \in {\cal U}$, then $\theta_c = \theta_a$ and if 
$n^t_c n^t_b \neq 0$ with $t \in {\cal U}$, then $\theta_c = \theta_b$ which yields the term $\theta_a \theta_b$ in the denominator. 
To obtain the last line, we have used the equalities $N^c_{ab}=N^{\bar{b}}_{a\bar{c}}=N^{\bar{c}}_{\bar{a}\bar{b}}$ and $N^c_{ab}=N^c_{ba}$.  
We now use \eqnref{eq.commuteRstrFus} to replace the terms in the round brackets and find
\begin{equation}\label{eq.MSSM.1}
[M, S]_{ab}= \frac{1}{D_\mA \theta_a \theta_b} \sum_{s\in\mT} \sum_{x \in \mA}  \theta_x d_x n_x^s \sum_{t\in\mU; r\in\mT} \left(n_a^t n_b^r  \tilde{N}_{rs}^t - n_b^t n_a^r \tilde{N}_{r\bar{s}}^t\right),
\end{equation}
\end{widetext}
where we have used the equality $n_x^s = n_{\bar{x}}^{\bar{s}}$ (the assumption that the restriction of $x$'s antiparticle $\bar{x}$ is the antiparticle of the restriction of $x$) to transfer the antiparticle on the $\tilde{N}^t_{r,\bar{s}}$.
We can now split up the  $r$ sum in Eq.~\eqref{eq.MSSM.1} into a sum over $\mU$ and a sum over $\mT/\mU$. For the first contribution, we have
\begin{equation}
\sum_{r,t\in\mU} \left(n_a^t n_b^r \tilde{N}_{rs}^t - n_b^t n_a^r \tilde{N}_{r\bar{s}}^t\right)
=
\sum_{r,t\in\mU} n_a^t n_b^r\left( \tilde{N}_{rs}^t - \tilde{N}_{t\bar{s}}^r\right)
\label{eq: intermediate step in M S commutator}
\end{equation}
by exchanging the labels $r$ and $t$ in the second term. Since $\tilde{N}_{t\bar{s}}^r=\tilde{N}_{\bar{r}\bar{s}}^{\bar{t}}=\tilde{N}^t_{rs}$, \eqnref{eq: intermediate step in M S commutator} vanishes identically. Thus, $r$ in \eqnref{eq.MSSM.1} can only take values in $\mT/\mU$.
By assumption, $\mU$ is a closed fusion category. This implies that no trivalent vertex with a single leg in $\mT/\mU$ exists in $\mT$. 
As a result, the $s$-sum in \eqnref{eq.MSSM.1} may only run over $\mT/\mU$. However, for $s\in\mT/\mU$ we can use the assumption \eqnref{a2} to find $\sum_{x\in\mA} \theta_x d_xn_x^s=0$ for the remaining terms in \eqnref{eq.MSSM.1}. We conclude that \eqnref{eq.MSSM.1} vanishes identically and thus $[S,M]=0$.

These equations are essential to the theory of condensation, as they establish that the condensation matrix $M_{ab}$ is a particular symmetry of the $S$ and $T$ modular matrices. While there exist other such symmetries, for example automorphisms  that are represented by permutation matrices, these matrices are not `completely positive' integer matrices i.e., they cannot be factorized as $nn^{\mathsf{T}}$ in terms of a nonnegative integer matrix $n$.

\section{The Modular Tensor Category after condensation}
\label{sec.MTC}

In the previous section, we identified a matrix $M$ which commutes with the modular matrices $S$ and $T$ of the $\mA$ theory.  In this section we prove a stronger pair of results, namely that
\begin{subequations}
\begin{eqnarray}
\label{eq.modinvS1}
n\tilde{S}&=& Sn, \\
\label{eq.modinvT1}
n\tilde{T}&=& Tn,
\end{eqnarray}
\end{subequations}
where $\tilde{S}$ and $\tilde{T}$ are the modular matrices of the $\mU$ theory and $n$ is the matrix of coefficients that enter the restriction and lifting maps, $(n)_{at}=n_a^t,~\forall~a\in\mA,~t\in\mU$. Our assumption \eqnref{a2} will be crucial for these proofs. The second equality~\eqnref{eq.modinvT1} is the statement that, component by component, whenever $n_a^t\neq 0$, $\theta_t=\theta_a$; this is true by recalling our definition of deconfined particles of the $\mU\ (\subset\mT)$ theory,  Eq.~(\ref{deconfined}). 

Before starting the proof of  the first equality~\eqnref{eq.modinvS1},  
we note the following equalities derived in Appendix~\ref{app.lemma} 
\begin{subequations}
\begin{equation}
\sum_{t\in\mU} \beta_t \beta_t^*=q^2 D_\mU^2
\label{eq: DU from beta}
\end{equation}
and
\begin{equation}
\sum_{t\in\mT} \beta_t \beta_t^*=D_\mA^2.
\label{eq: DA from beta}
\end{equation}
\end{subequations}
It then follows from assumption~\eqnref{a2}, that $\sum_{t\in\mU} \beta_t \beta_t^*=\sum_{t\in\mT} \beta_t \beta_t^*$ and thus $q^2 D_\mU^2=D_\mA^2$ or
\begin{equation}
q=D_\mA/D_\mU.
\label{eq: q relation}
\end{equation}

To prove \eqnref{eq.modinvS1}, we multiply \eqnref{eq.commuteRstrFus} by $\theta_a d_a$ and sum both sides over $a\in\mA$ to obtain
\begin{equation}
\begin{split}
\sum_{r,s\in\mT} \tilde{N}^t_{rs}n^s_b\beta_r
=&\, \sum_{a,c\in\mA} N^c_{ab}d_a\theta_a n^t_c\\
=&\, \theta_b D_\mA \sum_{c\in\mA}S_{bc}\theta_c n^t_c,
\end{split}
\label{eq: intermediate step in nS proof 0}
\end{equation}
where we have used the definition of $S$ from \eqnref{eq: def S} and that $\theta_a=\theta_{\bar{a}}$. For particles $t\in\mU$, we furthermore have
\begin{equation}
\sum_{r,s\in\mT} \tilde{N}_{rs}^t n_b^s \beta_r = \sum_{r,s \in \mU} \tilde{N}_{rs}^t n_b^s q \theta_r d_r,\qquad\forall t \in \mU,
\label{eq: intermediate step in nS proof}
\end{equation}
because 
(i) only $r\in\mU$ contributes to the sum (as $\beta_r=0$ if $r\in\mT/\mU$) and
(ii) only $s\in\mU$ contributes since $\mU$ is closed under fusion by assumption. We furthermore used \eqnref{a1} to rewrite the righthand side of \eqnref{eq: intermediate step in nS proof}. Since we assumed that $\mU$ forms a braided fusion category with its $\tilde{S}$ matrix, we use the usual definition of the $\tilde{S}$ matrix to write
\begin{equation}
\sum_{r,s\in\mU} \left(\tilde{N}_{rs}^t \theta_r d_r \right)n_b^s
=\sum_{s \in \mU}\tilde{S}_{st} D_\mU \theta_t \theta_s n_b^s.
\label{eq: intermediate step in nS proof 2}
\end{equation} 
Since $s,t \in \mU$, we furthermore have $\theta_s = \theta_b$ if $n^s_b\neq 0$ which allows us to combine \eqnref{eq: intermediate step in nS proof 0} and \eqnref{eq: intermediate step in nS proof 2} into
\begin{eqnarray}
\frac{1}{q}\frac{D_\mA}{D_\mU}\sum_{c\in\mA}S_{bc}\theta_c n_c^t=\sum_{s\in\mU}\tilde{S}_{st} n_b^s \theta_t.
\end{eqnarray} 
Since for all $n_c^t\neq 0$, we have $\theta_c=\theta_t$ ($t\in\mU$) and using \eqnref{eq: q relation}, this expression reduces to
\begin{eqnarray}
\sum_{c\in\mA}S_{bc}n_c^t=\sum_{s\in\mU}\tilde{S}_{st} n_b^s.
\label{eq: s-n}
\end{eqnarray}

We have thus proven \eqnref{eq.modinvS1} and \eqnref{eq.modinvT1} within our algebraic formulation of the condensation transition. These two equations have a well known parallel in the study of chiral algebra extensions, which we detail in Appendix~\ref{sec.chiralextension}. As a side remark, let us derive a consequence of \eqnref{eq: q relation}, namely
\begin{equation}
D_\mA > D_\mU, 
\label{eq: D theorem}
\end{equation}
which follows  from the fact that the embedding  dimension $q = \sum_{a \in\mA} d_a n^\varphi_a > 1$ and $q=1$ if no condensation is happening. This is always true even if the assumption Eq.~\eqref{a2} is not used. If  Eq.~\eqref{a2} is used, then Eqs.~\eqref{eq: DU from beta} and \eqref{eq: DA from beta} imply $D_\mA= q D_\mU$. By analogy with the Zamolodchikov $c$-theorem  of Ref.~\onlinecite{Zamo} one can  call this result  the ${D}$-theorem which can be interpreted as the disappearance of some anyons upon condensation. There is a stronger connection between this result and the $g$-theorem, according to which the Affleck-Ludwig boundary entropy of  an open conformal system decreases under the renormalization group flow of the boundary as long as the bulk theory remains critical throughout the flow.\cite{AL91,FK,GMS}  (This situation is, however, distinct from the case of a condensation transition in which the bulk is only critical at the transition.)  The boundary entropy is in turn related to the  quantum dimension  of the primary field  that characterizes the boundary condition,  which suggests a relation between the $g$-theorem and \eqnref{eq: D theorem}.

\section{Simple currents}
\label{sec.simplecurrent}

Simple currents are abelian anyons that, when raised to a certain power by fusion, equal the identity, see Refs.~\onlinecite{SY89,SY90,I90,GS1,GS2,KS94,FFS96}. The precise definition follows below. In this section, we consider a condensate that is composed of simple currents only. In this situation, we can prove that \eqnref{a2}, i.e., $\beta_t=0,~\forall t\in\mT/\mU$, follows from the assumptions in \secref{sec.def}.

\subsection{Introduction to simple currents}

\begin{subequations}
There are several equivalent definitions of simple currents in the context of rational conformal field theory (RCFT). First, a simple current is a primary field $J$ that has a unique fusion channel with any other primary field of the RCFT 
\begin{eqnarray}
J \times \phi = \phi',~~\forall \phi.	\label{s1}
\end{eqnarray}
A second definition is that a simple current is a primary field $J$ that when fused with its antiparticle or conjugate field  $\bar{J}$ only fuses to the identity (see Ref.~\onlinecite{I90})
\begin{eqnarray}
J \times \bar{J} =  1	. 	\label{sb1}
\end{eqnarray}
A third definition is that the quantum dimension of $J$ is 1,
\begin{eqnarray}
d_J = 1 . 	\label{sbc1}
\end{eqnarray}
One can show that all these definitions are equivalent.~\cite{SY90}
\end{subequations}

Given two simple currents $J_1$ and $J_2$, their fusion product $J_1 J_2$ is also a simple current. The number of primary fields of a RCFT is finite, therefore each simple current $J$ has an associated integer $N$ such that $J^N = 1$ by using Eq.~\eqref{s1}. The smallest integer $N>0$ with this property is called the order of $J$. A simple current $J$ generates a set of simple currents $\{ J^m |  m=0, 1, \dots, N-1\}$, which is isomorphic to the abelian group $\Zmath_N$. A RCFT may contain simple currents generated by more that one primary field. The collection of all of them form an Abelian group which is  isomorphic to the product $\Zmath_{N_1} \times \dots \times  \Zmath_{N_r}$. One can choose a basis of simple currents such that $N_i$ are of the form $p_i^{n_i}, \, n_i \in \Zmath$, with $p_i$ a prime number. This is the fundamental theorem of finite abelian groups. 

As an example, consider the RCFT constructed from 
the Kac-Moody algebra SU(2)$_k$. The primary fields are denoted by $\phi_\ell$ where $\ell \; =0,1,  \dots, k$
is twice the topological spin. The field $\phi_k$ is a simple current because its  fusion rule is $\phi_k \times \phi_\ell  = \phi_{k - \ell}$.  
Indeed $\phi_k$ is the only non-trivial simple current, and satisfies $\phi_k \times \phi_k = \phi_0 = 1$. The simple currents form a $\mathbb{Z}_2=\{\phi_0,\phi_k\}$ subcategory. 

When acting on a primary field $\phi$, $J$ generates an orbit formed by the fields $J^n \phi$ 
\begin{eqnarray}
[\phi] = \{ \phi, J \phi , J^2 \phi, \dots, J^{ d-1} \phi \}, \quad J^d \phi = \phi.		\label{sb4}
\end{eqnarray}
Here, $d$ is the smallest positive integer such that $J^d \phi = \phi$. The orbit~\eqnref{sb4} is denoted by a representative field $\phi$ but one can choose another field belonging to the orbit. In general $d$ need not equal $N$, the order of the current $J$, but $d$ must divide $N$. In the example of SU(2)$_k$, if $k$ is odd, all the orbits
have two elements, ($d=N=2$), while for $k$ even, $\phi_k\times \phi_{\frac{k}{2}} = \phi_{\frac{k}{2}}$ for the action of $\phi_k$ and so the orbit has only one element, $\phi_{\frac{k}{2}}$. Generally, we will simply call the anyon  a ``fixed point" when it is invariant under fusion with a simple current, or equivalently, if its orbit contains only the anyon itself. In the SU(2)$_k$ ($k$ even) example, $\phi_{\frac{k}{2}}$ is a fixed point under the fusion with $\phi_k$.
As we shall see below, the existence of fixed points is  crucial for the construction of the condensed theory.

\subsection{Simple current condensation}\label{ssec:SCC}

We consider a condensation transition, where the condensate consists only of the set of bosonic simple currents, generated by $n$ simple currents $J_1,\ldots, J_n$ with orders $N_1,\ldots, N_n$. Any anyon in the condensate can thus be represented as $J_1^{i_1}\ldots J_n^{i_n}$, where $i_l=0,\ldots, N_l-1$ and $l=1,\ldots, n$ (note that the fusion product of simple currents is unique.) We use the shorthand notation  $\bs{i}=(i_1,\ldots, i_n)$ and 
\begin{equation}
J_{\bs{i}}:=J_1^{i_1}\cdots J_n^{i_n}.
\end{equation}

The initial theory might have simple currents which are not bosons. We do not consider these, as they cannot condense. We consider the group generated by the powers of all the \emph{bosonic} simple currents, which is sometimes called the bosonic center $\mathcal{C}$ of the RCFT. Powers of a condensed bosonic simple current, or the products of different condensed bosonic simple currents are also bosonic and condensed. To see this, examine the $J_{\bs{i}},J_{\bs{j}}$ component of Eq.\eqref{eq.commuteRstrFus}, where $J_{\bs{i}},J_{\bs{j}}$ are assumed to be condensed.  Recalling that $J_{\bs{i}+\bs{j}}:=J_{\bs{i}}\times J_{\bs{j}}$ is automatically a simple current (see above), and making use of the fact that condensed simple currents like $J_{\bs{i}}$ have quantum dimension $1$ so that $n^{t}_{J_{\bs{i}}}=\delta^t_{\varphi}$, we find
\begin{equation}
1=n^\varphi_{J_{\bs{i}}} n^\varphi_{J_{\bs{j}}}=N_{J_{\bs{i}},J_{\bs{j}}}^{J_{\bs{i}+\bs{j}}} n^\varphi_{J_{\bs{i}+\bs{j}}}=n^\varphi_{J_{\bs{i}+\bs{j}}}.
\end{equation}
As a result  $n^\varphi_{J_{\bs{i}+\bs{j}}}=1$, indicating that $J_{\bs{i}+\bs{j}}$ restricts solely to the vacuum, so it must be a boson. Therefore the product of any two condensed simple current is a condensed (hence bosonic) simple current.

As an aside, we note that for general bosonic currents which are not necessarily condensed: (i) as before, any power of such a simple current is a  bosonic simple current; (ii) however, the product of two such  bosonic simple currents does not have to be bosonic. For example, in the toric code that we will discuss in detail in Sec.~\ref{sec.examples}, $e$ and $m$ are bosonic simple currents while their product $f$ is actually a fermion (which cannot condense). To prove (i), one can use the symmetry $S_{ab}=S_{a\bar{b}}^*$ of the $S$ matrix and the fact that for any anyon $\theta_a=\theta_{\bar{a}}$. Choosing $a=b=J$ with $J$ a simple current gives
\begin{equation}
\frac{\theta_{J^2}}{\theta_{J}\theta_{J}}=\frac{\theta_{1}}{\theta^*_{J}\theta^*_{\bar{J}}}.
\end{equation} 
If $J$ is bosonic, then so is $\bar{J}$ and the above implies $J^2$ is also a bosonic simple current, i.e., $\theta_{J^2}=1$. This argument can be iterated by assuming that up to some $n_0$ all $J^n$, $n=1,\ldots, n_0$, are bosonic (then so are all $J^{N-n}$, $n=1,\ldots, n_0$, with $N$ the order of $J$). Solving the equality $S_{J,J^{n_0}}=S_{J,J^{N-n_0}}^*$ for $\theta_{J^{n_0+1}}$ yields that $J^{n_0+1}$ is also bosonic.

\subsubsection{Vafa's theorem}
We first aim to find information about the topological spins of some of the particles in the theory by analyzing the implications of Vafa's theorem~\cite{Vafa}
\begin{equation}
\begin{split}
&\prod_p \left(\frac{\theta_p}{\theta_x \theta_y }\right)^{N_{xy}^p N_{pz}^u}  \prod_q \left(\frac{\theta_q}{\theta_x \theta_z} \right)^{N_{yq}^u N_{xz}^q} 
\\
&\qquad= \prod_r \left(\frac{\theta_u}{\theta_x \theta_r} \right)^{N_{yz}^r N_{xr}^u} 
\end{split}
\end{equation}
$\forall x,y,z,u$.
For the case of the simple current condensate, we pick a particle $x=a$, a particle  $y= J_{\bs{i}}$ and  a particle $z=J_{\bs{j}}$. Note that $a$ can be any particle in the $\mA$ theory, not necessarily a simple current. This choice of the anyons uniquely fixes all other anyons in the equation ($p = a\times J_{\bs{i}}$, $u=  a\times J_{\bs{i}+\bs{j}}$,  $q=a\times J_{\bs{j}}$, $r=  J_{\bs{i}+\bs{j}}$). Using the fact that the simple currents and their powers are all bosons, Vafa's theorem gives
\begin{eqnarray}\label{eq:character of simple current}
\frac{ \theta_{  a\times J_{\bs{i}}}}{\theta_a} 
\frac{\theta_{ a\times J_{\bs{j}}  }}{\theta_a} 
 = 
 \frac{ \theta_{    a\times J_{\bs{i}+\bs{j}}}}{\theta_a}.
\end{eqnarray}
This equation implies that the fractions $\theta_{a\times J_{\bs{i}}}/\theta_a$ are irreducible characters of the group $\Zmath_{N_1} \otimes \Zmath_{N_2} \otimes \ldots \otimes \Zmath_{N_n}$, the bosonic center of RCFT which condenses. The one-dimensional characters of this group can be written as
\begin{eqnarray}
\frac{ \theta_{  a\times J_{\bs{i}}}}{\theta_a} 
=\omega_1^{i_1}\omega_2^{i_2}...\omega_n^{i_n},
\label{eq: omegas} 
\end{eqnarray}
where $\omega_i$'s satisfy $\omega_i^{N_i}=1$. Also note that the $\omega_i$'s secretly depend on the subindex $a$. There are two cases:

\begin{subequations}
\begin{eqnarray}
&&
\textit{Case 1}:\qquad\qquad \theta_{  a\times J_{\bs{i}}}/\theta_a
=1,\hspace{3cm}\\
&&
\textit{Case 2}:\qquad\qquad \theta_{  a\times J_{\bs{i}}}/\theta_a
 \ne 1.
\end{eqnarray} 
\end{subequations}

In the latter case, if particles $a$ and ${a \times  J_{\bs{i}}}$ restrict to the same particle $t\in\mT$ then this particle is confined (as  $\theta_{a \times  J_{\bs{i}}} \ne \theta_a$). Moreover, from the orthogonality of characters \eqnref{eq: omegas}   we know that in this case
\begin{equation}
\begin{split}
\sum_{i_1, i_2, \ldots, i_n}^{N_1, N_2, \ldots, N_n} 
\frac{ \theta_{  a\times J_{\bs{i}}}}{\theta_a} 
&= \sum_{i_1=0}^{N_1-1}\omega_1^{i_1}\sum_{i_2=0}^{N_2-1}\omega_2^{i_2}...\sum_{i_n=0}^{N_n-1}\omega_n^{i_n} 
\\
&=0 .
\label{svc8}
\end{split}
\end{equation}
This happens when at least one $\omega_{i_l}$ is not equal to $1$.

\subsubsection{Condensation}

Without loss of generality, we will assume $J_1, J_2,\ldots, J_n$ condense. If only a subset of the simple currents condense then the same analysis applies to just the bosons that condense (the others factor out). Since $d_{J_1} = \ldots = d_{J_n}=1$, the bosons restrict only to the new vacuum $\varphi$ with  coefficients unity
\begin{eqnarray}
n_{J_1}^\varphi =\ldots =n_{J_n}^\varphi =1
\end{eqnarray}
and do not split. Using the reasoning in \secref{ssec:SCC} it follows that all products of these simple currents also condense -- indeed, all bosonic simple currents $J_{\bs{i}}$ condense.

We will now proceed to prove a few crucial lemmas for any $a,b\in \mathcal{A}$: (i) $n^t_a=n^t_{a\times J_{\bs{i}}},~\forall i$, for all $t \in \mathcal{T}$ and (ii) $\sum_t n^t_a n^t_b\neq 0$ if and only if $b= a\times J_{\bs{j}}$ for some $\bs{j}$. 

(i) is easily proved by examining the $b= J_{\bs{i}}$ component of \eqnref{eq.commuteRstrFus}. To show (ii) we examine the $t=\varphi$ component of \eqnref{eq.commuteRstrFus} and note that all bosonic simple currents condense giving
\begin{equation}
\begin{split}
\sum_{t\in\mT}  n_a^t n_b^t 
&= \sum_{i_1, i_2, \ldots i_n}^{N_1, N_2, \ldots, N_n} 
N_{a, J_{\bs{i}}}^b   
n_{J_{\bs{i}}}^\varphi  
\\
&= 
\sum_{i_1, i_2, \ldots, i_n}^{N_1, N_2, \ldots, N_n} \delta_{b , a \times J_{\bs{i}}}  .
\label{scv10}
\end{split}
\end{equation}
for any $a,b\in \mathcal{A}$. To prove (ii), note that:
\begin{itemize}
\item If $b \ne a \times J_{\bs{j}}$ for all  $\bs{j}$, then $\sum_{t \in\mT}  n_a^t n_b^t =0$ and particles $a,b$ do not have any common restrictions. 
Let us write this  result as 
\begin{eqnarray}
{\rm If} \; b \notin [a]  \  \Longrightarrow \ n^t_a n^t_b = 0,\ \forall t, 
\end{eqnarray}
where $[a] = \{ J_{\bs{j}} a ,  J_{\bs {j}}  \in {\cal C} \}$ is the orbit obtained  acting on $a$  with  all the bosonic simple currents. 
\item If $b = a \times J_{\bs{j}}$ for some $\bs{j}$ then 
\begin{equation}
\begin{split}
 \sum_{t\in\mT}  n_a^t n_{a \times J_{\bs{j}}}^t
=& \sum_{i_1, i_2, \ldots i_n}^{N_1, N_2, \ldots, N_n} \delta_{ a \times J_{\bs{j}} , a \times J_{\bs{i}}}	\\
=& R_a \in \mathbb{Z}_{+}.
\end{split}
\end{equation}
But from (i), $ n_{a \times J_{\bs{j}}}^t = n^t_a$, so the LHS of this equation is positive. Hence $R_a>0$. For example, for $n=1$, $R_a=N_1/d$, with $d$ defined in Eq.~\eqref{sb4}.
\end{itemize}
Hence we have proved (ii). From (ii) we know that if $a \text{ and } b$ are in the lift of $t$ then $b=a \times J_{\bs{j}}$ for some $\bs{j}$. On the other hand from (i), if $a$ is in the lift of $t$, so is $a\times  J_{\bs{j}}$. Hence $t$ is deconfined iff $\theta_{a} = \theta_{a \times J_{\bs{j}}}$ for all $\bs{j}$, where $a$ is any particle in the lift of $t$. In other words, given an $a\in \mathcal{A}$, the character $ \theta_{a \times J_{\bs{j}}}/\theta_{a}\neq 1 \text{ for some } \bs{j}$ iff $a$ restricts only to confined particles. 

Let us now prove the assumption~\eqref{a2}. We first  multiply \eqnref{scv10} by $d_b \theta_b$ and sum over all particles $b$ in the $\mA$ theory to obtain
\begin{equation}
\begin{split}
\sum_{t\in\mT} \beta_t n_a^t &= \sum_{i_1, i_2, \ldots, i_n}^{N_1, N_2, \ldots, N_n} d_{ a \times J_{\bs{i}}} \theta_{ a \times J_{\bs{i}}} \\
&= d_a \theta_a \sum_{i_1, i_2, \ldots, i_n}^{N_1, N_2, \ldots, N_n} 
\frac{ \theta_{  a\times J_{\bs{i}}}}{\theta_a} ,
\end{split}
\end{equation}
where we have used the fact that the quantum dimension of any product of a particle with simple currents remains the same. Now if $\theta_{  a\times J_{\bs{i}}}/\theta_a$ is not the identity character of the trivial representation, then the particle $a$ restricts only to confined particles and we have from \eqnref{svc8}
\begin{equation}
\sum_{t\in\mT} \beta_t n_a^t  = \sum_{t\in\mT/\mU} \beta_t n_a^t =0.
\end{equation}
In fact, the second equality  holds even if $\theta_{  a\times J_{\bs{i}}}/\theta_a=1$ for all $\bs{j}$, because in that case $n_a^t=0\forall t\in \mT/\mU$ -- as a result, the second equality holds for all $a$. Multiplying the second equality by $\theta_a^* d_a$ and summing over all particles $a$ we obtain
\begin{eqnarray}
0
=\sum_{t\in\mT/\mU} \beta_t \sum_{a\in\mA} \theta_a^* d_a n_a^t 
= \sum_{t\in\mT/\mU} \beta_t \beta_t^*.
\end{eqnarray}
The unique solution is $\beta_t=0$ for $t$ confined, coinciding with our assumption~\eqref{a2}. 

\section{One confined particle theories}
\label{sec: One confined particle theories}

In this section we study a simple boson condensation with just one confined particle $t_0$ in the $\mT$ theory. Furthermore, assume that the confined particle $t_0$ has only two lifts $a_1$ and $a_2$ with lifting coefficients both $1$, i.e.,
\begin{eqnarray}
&& n^{t_0}_{a_1}=n^{t_0}_{a_2}=1,	\\
&& \text{otherwise}~n^{t_0}_a=0, \forall a\neq a_1,a_2.
\end{eqnarray}

With these assumptions, we can prove that the condensate has only one boson besides vacuum, and this condensed boson has quantum dimension $1$. 
This implies that the boson is a simple current, so the results of the previous section imply $\beta_{t_0}=0$. However, we choose to prove this equation through another method which gives more information about bosonic condensation theories with only one confined particle. 
Further we find that $d_{t_0}=d_{a_1}=d_{a_2}$, which means that $a_1$ and $a_2$ only restrict to one particle $t_0$ in the $\mT$ theory, with no other particles in $\mT$. Finally, in this special one-confined particle case, we prove that $\beta_{t_0}:=\sum_{a\in\mA} n_a^{t_0} d_a \theta_a=0$ which clearly support the assumption we used in previous sections. We give the detailed proof in Appendix \ref{app.beta}.

\section{Formalism and implementation}
\label{sec.examples}

We now present an algorithmic prescription, which can be implemented on a computer, and which strongly contrains the possible condensation transitions starting from a TQFT with given modular matrices $S$ and $T$. We  then apply this procedure to several example TQFTs. The algorithm is performed in 3 steps:
\begin{enumerate}
\item 
Search for the symmetric matrices $M$ with nonnegative integer entries and $M_{1,1}=1$  satisfying
\begin{equation}
[M,S]=[M,T]=0.
\label{eq: M, T commutator}
\end{equation}
\item
For each $M$, find all nonnegative integer rectangular matrices $n$ such that $M=nn^{\mathsf{T}}$.
\item
For each $M$ and $n$, find the putative  modular matrices $\tilde{S}$ and $\tilde{T}$ of the TQFT after condensation by solving 
\begin{equation}
Sn=n\tilde{S}\  \text{and} \ Tn=n\tilde{T}.
\label{eq: equation for S matrix}
\end{equation}
One subtlety is that we need to make sure that the $\tilde{S}$, $\tilde{T}$ matrices we obtain are valid. In this paper, we use the necessary conditions for a valid $S$ matrix: it should be symmetric, unitary, and it should generate non-negative fusion coefficients by Verlinde formula. These are always satisfied if $\mathcal{U}$ is a MTC.
\end{enumerate}
This algorithm sidesteps the discussion of the theory $\mT$ that contains confined anyons and directly yields the resulting MTC $\mU$ formed by the remaining deconfined anyons. The algorithm provides all condensation solutions of theory $\mA$. Another algorithm which does not sidestep $\mT$ is to (1) build the matrix $M'$ as in the bracket of Eq.~\eqref{eq: M equation for T}; (2) factorize it in $n_a^t$; (3) keep only the deconfined particle $t$'s; and then apply step (3) and Eq.~\eqref{eq: equation for S matrix}.
Whether the two theories are identical hinges on Eq.~\eqref{a2}, which we assume to be true. 
We now address the above steps one by one.

\subsection{Solutions for $M$}

Since $T$ is diagonal, the equation $[M,T]=0$ is satisfied if and only if $M$ is a block diagonal matrix with nonzero off-diagonal entries only between particles with the same topological spin. Imposing this block structure, we can solve $[M,S]=0$, imposing that 
\begin{equation}
\begin{split}
1.~M_{ab}=& M_{ba}\geq0,\qquad M\in\mathbb{Z},	\\
2.~M_{11}=& 1.
\end{split}
\label{eq: condition on M}
\end{equation}
The second condition ensures that the $\mA$ vacuum restricts to the  vacuum $\varphi$ of $\mU$. 
In this case, the first row (or column) of $M$ is equal to the first column of $n$ and describes the particles that condense into the vacuum $n^\varphi_a$. (From this, it is also clear that only solutions with $M_{1a}\leq d_a$ can lead to a valid theory.)
 
With conditions 1 and 2 in Eq.~\eqref{eq: condition on M} in place, we obtain two types of solutions for $M$, which we call automorphisms and condensations, aside from the trivial solution $M=\openone$.

Automorphisms are defined by a fully-ranked matrix $M$ satisfying
\begin{equation}
\sum_a M_{ab}=1\qquad \forall b.
\end{equation}
They satisfy $M^2=\openone$ because of the following reasons: Since $\sum_a M_{ab}=1$ and all entries of $M$ can only be nonnegative integers, for any $b\in\mA$, there is only one corresponding particle $b'$, such as $M_{b'b}=1$. Further, $M_{ab}=0,~\forall a\neq b'$ and $M$ is fully-ranked. As a result, if $a \neq b$ then $a' \neq b'$. Hence $(M^2)_{ab}=\sum_c M_{ac}M_{cb}=\sum_c M_{ac}M_{bc}=\delta_{ab}$. 
An automorphism $M$ is thus a permutation matrix of order two -- it is a symmetry of the $S,T$ data under relabeling of particles. All automorphisms of $\mA$ form a group under matrix multiplication, which is used to construct ``topological symmetry group" in the presence of a global symmetry.~\cite{BBCW14} Automorphism, however, still exists even when any other symmetries (e.g. $U(1)$ charge conservation), are broken. 

On the other hand, solutions $M$ that correspond to a condensation have $M_{1a}\neq\delta_{a,1}$ for some $a$, implying that at least one other boson besides the vacuum restricts to the new vacuum. All the condensations can be superimposed with any of the automorphisms, yielding a potentially different condensation. In other words, two condensations can be related via a permutation of $\mA$ by multiplying the $M$ matrix of one condensation from both sides with the $M$ matrix of the automorphism -- we will see an example of this below for the toric code TQFT.

We can prove that any $M$ that satisfies Eq.~\eqref{eq: M, T commutator} and conditions 1 and 2 in Eq.~\eqref{eq: condition on M} is either an automorphism or a condensation as follows. We first assume that $M$ is not a condensation solution, that is, the first row and column of $M$ are all zeros ($M_{1a}=M_{a1}=0,~\forall a\neq 1$) except $M_{11}=1$ . We show that $M$ must be an automorphism in this case.
From
\begin{equation}
\sum_bM_{ab}S_{bc}=\sum_bS_{ab}M_{bc}
\end{equation}
we have for $c=1$
\begin{equation}
\sum_bM_{ab}S_{b1}=S_{a1}
\ \Rightarrow\ 
\sum_bM_{ab}d_{b}=d_{a}.
\label{eq: Md eigenvector}
\end{equation}
Thus, $d_a$ is a strictly positive eigenvector of $M$ with eigenvalue $1$. Since every $M_{ab}$ is integer and larger or equal to zero, \eqnref{eq: Md eigenvector} can only hold if
\begin{equation}
f_a \equiv \sum_bM_{ab}\geq1.
\end{equation}
On the other hand, summing \eqnref{eq: Md eigenvector} over $a$, and using $M=M^\mathsf{T}$, gives
\begin{equation}
\sum_{b}f_b d_{b}=\sum_ad_{a}.
\end{equation}
Again, since $f_a \geq 1$ and the $d_a$ are strictly positive this equation can only be satisfied if
\begin{equation}
f_a \equiv \sum_bM_{ab}=1,
\end{equation}
which, together with the fact that $M$ is symmetric, implies that $M$ has to be an automorphism (a permutation matrix).

\begin{subequations}
\label{eq: Toric code}
Let us illustrate how automorphism and condensation solutions for $M$ arise from condition~\eqref{eq: M, T commutator} for the example of the toric code (TC) TQFT. It contains the anyons 1, $e$, $m$, $f$ and has the modular matrices
\begin{equation}
S_{\mathrm{TC}}=\frac{1}{2}
\begin{pmatrix}
1&1&1&1\\
1&1&-1&-1\\
1&-1&1&-1\\
1&-1&-1&1
\end{pmatrix}
\end{equation}
and
\begin{equation}
T_{\mathrm{TC}}=\mathrm{diag}
(1,1,1,-1).
\end{equation}
\end{subequations}
It admits three nontrivial solutions to \eqnref{eq: M, T commutator}, one automorphism
\begin{equation}
M^{(1)}=
\begin{pmatrix}
1&0&0&0\\
0&0&1&0\\
0&1&0&0\\
0&0&0&1
\end{pmatrix}
\end{equation}
that exchanges the $e$ and the $m$ particles and two condensations
\begin{equation}
M^{(2)}=
\begin{pmatrix}
1&1&0&0\\
1&1&0&0\\
0&0&0&0\\
0&0&0&0
\end{pmatrix},
\quad
M^{(3)}=
\begin{pmatrix}
1&0&1&0\\
0&0&0&0\\
1&0&1&0\\
0&0&0&0
\end{pmatrix},
\end{equation}
of either the $e$ or the $m$ boson. They are related by the automorphism $M^{(2)}=M^{(1)}M^{(3)}M^{(1)}$ [note that $(M^{(1)})^{-1}=M^{(1)}$].

\subsection{Solutions for $n$}

Next we solve for the integer matrix $n^t_a\geq0$, where $t$ labels the {\emph{deconfined}} particles in the MTC $\mU$. It is possible that multiple solutions $n$ exist for a given $M$. However, for some solutions, it still might not be possible to find a valid condensed MTC: please refer to our Appendix~\ref{sec: 4 layers Ising} for an example of 4-layer Ising model condensation. In that example, we obtain unitary $S$ and  $T$ matrices,  
but they do not correspond, via Verlinde's formula, to integer fusion coefficients. 

An
efficient first step in solving for $n$ is to realize that any column of $M$ that only contains zeros and ones is equal to a column in $n$. While the matrix $M$ may contain several columns with only zeros and ones that are equal, they all correspond to only a single column in $n$ (there are no duplicate columns in $n$). 
After removing from $M$ all rows and columns that contain only zeros and ones, an actual factorization routine can be used on the remaining sub-block of the $M$ matrix. (As we will discuss for an example below, the factorization does not always yield a unique solution for $n$ in this case.) In the situations we have encountered, this part of the algorithm is not limited by computational power. In the particularly simple toric code example, deleting duplicate columns directly yields the solution
\begin{equation}
M^{(2)}=nn^{\mathsf{T}},
\qquad
n^{\mathsf{T}}=(1,1,0,0).
\end{equation}
There is only one particle in the new theory, the vacuum. Thus, condensation of either the $e$ or the $m$ particle in the toric code yields the trivial TQFT.

As a less trivial example, consider a bilayer of Ising TQFTs. Each layer contains the anyon types $1,\ \sigma,\ \psi$ with modular matrices
\begin{equation}
S_{\mathrm{I}}=\frac{1}{2}
\begin{pmatrix}
1&\sqrt{2}&1\\
\sqrt{2}&0&-\sqrt{2}\\
1&-\sqrt{2}&1
\end{pmatrix},
\quad
T_{\mathrm{I}}=
\mathrm{diag}(1,e^{\mathrm{i}\pi/8},-1).
\label{eq: Ising}
\end{equation}
The bilayer $S$ and $T$ matrices are direct products 
$S_{\mathrm{I}^{(2)}}=S_{\mathrm{I}}\otimes S_{\mathrm{I}}$, 
$T_{\mathrm{I}^{(2)}}=T_{\mathrm{I}}\otimes T_{\mathrm{I}}$,
and the theory supports 9 particle types which we denote 
$11,\ 1\sigma,\ 1\psi, \ \sigma1,\ \sigma\sigma,\ \sigma\psi,\
\psi1, \ \psi\sigma,\ \psi\psi$, where $11$ is the vacuum.
There is only one nontrivial solution for $M$, which reads in this basis
\begin{equation}
M=
\begin{pmatrix}
1&0&0&0&0&0&0&0&1\\
0&0&0&0&0&0&0&0&0\\
0&0&1&0&0&0&1&0&0\\
0&0&0&0&0&0&0&0&0\\
0&0&0&0&2&0&0&0&0\\
0&0&0&0&0&0&0&0&0\\
0&0&1&0&0&0&1&0&0\\
0&0&0&0&0&0&0&0&0\\
1&0&0&0&0&0&0&0&1
\end{pmatrix}.
\end{equation}
It is straightforward to obtain the unique solution $n$ that yields $M=nn^{\mathsf{T}}$
\begin{equation}
n^{\mathsf{T}}=
\begin{pmatrix}
1&0&0&0&0&0&0&0&1\\
0&0&0&0&1&0&0&0&0\\
0&0&0&0&1&0&0&0&0\\
0&0&1&0&0&0&1&0&0
\end{pmatrix},
\end{equation}
which shows that this describes the condensation of the $\psi\psi$ particle. In this process, the $\sigma\sigma$ particle (which has quantum dimension 2) splits into two particles of quantum dimension 1 
and both $1\psi$, $\psi1$ restrict to the same particle. All other particles, except for the vacuum, become confined. 

There exists $M$ that solve Eq.~\eqref{eq: M, T commutator}, but cannot be decomposed as
$M=nn^{\mathsf{T}}$ with a nonnegative integer matrix $n$. 
Some of them still admit an interpretation in terms of a condensation in the following sense. If the MTC $\mU$ that is obtained from a condensation with matrix $M=nn^{\mathsf{T}}$ has an automorphism symmetry $\tilde{P}$, that is equal to its transpose $\tilde{P}=\tilde{P}^{\mathsf{T}}$, then $\tilde{M}=n\tilde{P}n^{\mathsf{T}}$ is also a symmetric matrix that solves Eq.~\eqref{eq: M, T commutator}.
For instance, one necessary condition for a decomposition $M=nn^T$ to be possible is that $M_{aa}+M_{bb}\geq 2 M_{ab}$. If the matrix elements of $M$ do not satisfy the triangle equation $M_{aa}+M_{bb}\geq 2 M_{ab}$, then $M=n\tilde{P}n^T$ might be possible instead.

If the TQFT corresponds to a CFT, the possible forms of matrices $M$ that solve Eq.~\eqref{eq: M, T commutator} are understood with the help of the ``naturality theorem"  by Moore and Seiberg~\cite{MS89,MS89b}.
This theorem implies that all $M$ that solve Eq.~\eqref{eq: M, T commutator} in a CFT are either automorphisms of $\mA$,  condensations of the form $M=nn^{\mathsf{T}}$, or of the form $M=n\tilde{P}n^{\mathsf{T}}$, with $\tilde{P}$ an automorphism of $\mU$. As a corollary, we then conclude that for any solution to Eq.~\eqref{eq: M, T commutator} of the from $M^{(2)}=n\tilde{P}n^{\mathsf{T}}$, there is another solution $M^{(1)}=nn^{\mathsf{T}}$, with the same $n$, since the identity mass matrix of $\mU$ always exists.
For the purpose of studying condensations, we thus focused on matrices $M$ that admit the decomposition $M=nn^{\mathsf{T}}$ throughout our analysis. If we relaxed this constraint to also include $M=n\tilde{P}n^{\mathsf{T}}$, assumptions such as $n_a^t=n_{\bar{a}}^{\bar{t}}$ would not be justified anymore.
We discuss the interpretation of condensation transitions for CFTs in Appendix~\ref{sec.chiralextension} and relate it to the ``naturality theorem". Subsequently, in Appendix~\ref{app: SU2 condensation} we give an example of condensation transitions in SU$(2)_{16}$, for which two solutions $M=n\tilde{P}n^{\mathsf{T}}$ and $\tilde{M}=nn^{\mathsf{T}}$ to Eq.~\eqref{eq: M, T commutator} exist.

The decomposition $M=nn^{\mathsf{T}}$ is generally not unique. For example, if $M_{aa}=4$ for some particle with quantum dimension $4$ or larger, it can either split in 4 particles with $n_a^t=1$ for each or restrict to one particle with $n_a^t=2$ (this issue was discussed previously in Sec.~\ref{sec.simplecurrent}). 
However, in all examples we studied, at most one of all possible decompositions of $M$ lead to a consistent TQFT with valid solutions for $\tilde{S}$ and $\tilde{T}$. 
Thus, the uniqueness of this step in the condensation is an open question.
We mentioned that factorizing $M=nn^{\mathsf{T}}$ is a well-known problem in the field of completely positive matrices. In our cases, the factorization happens over the ring of positive integers. This problem is known to be NP-hard. With the exception of small dimension matrices, it has not yet been solved. Some outstanding questions are the characterization of when a matrix $M$ is completely positive (sufficient and necessary condition), as well as what is the minimal number of rows in $n$ (called CP rank), which is translated in our case to the minimal number of particles in $\mU$ that can be obtained.

\subsection{The modular matrices of the new theory}
Having obtained the matrix $n$, we now solve the equations
\begin{equation}
Sn=n\tilde{S},\qquad
Tn=n\tilde{T}
\label{eq: nS nT equations}
\end{equation}
for $\tilde{S}$ and $\tilde{T}$.
These equations can have spurious solutions unless we impose a list of additional constraints. For modular theories, these constraints are
\begin{itemize}
\item $\tilde{S}^\dagger=\tilde{S}^{-1}$,
\item $\tilde{S}^2=\Theta(\tilde{S}\tilde{T})^3=\tilde{C}$, where $\tilde{C}$ is a permutation matrix that squares to the identity and $\Theta=e^{-\mathrm{i}\pi c/4}$ with $c$ the chiral central charge of $\mA$, which we can prove remains unchanged (mod 8) during condensation.
\item $\tilde{T}$ is a diagonal matrix with complex phases on the diagonal,
\item the fusion coefficients obtained from the Verlinde formula 
\begin{equation}
\tilde{N}_t=\tilde{S}\tilde{D}_t\tilde{S}^{-1},
\label{eq: new fusion coefficients}
\end{equation}
 with $(\tilde{D}_t)_{rs}=\delta_{r,s}\tilde{S}_{tr}/\tilde{S}_{1r}$ have to be nonnegative integers.
\end{itemize}
We do not prove that any solution that obeys the above list of conditions is indeed a valid MTC $\mU$. However, any allowed condensation will be a solution to these conditions. Therefore, if we do not find a solution for a given MTC $\mA$, we can conclude that no condensation transition to a modular $\mU$ theory out of $\mA$  exists (we will discuss a nontrivial example for this situation in Sec.~\ref{sec: Fib}).

For the example of the double layer Ising theory, we have for 
$S_{\mathrm{I}^{(2)}} n = n \tilde{S}$  (skipping columns of zeros, which correspond to the confined particles) 
\begin{equation}
\begin{pmatrix}
\frac{1}{2}&\frac{1}{2}&1&\frac{1}{2}&\frac{1}{2}\\
\frac{1}{2}&-\frac{1}{2}&0&-\frac{1}{2}&\frac{1}{2}\\
\frac{1}{2}&-\frac{1}{2}&0&-\frac{1}{2}&\frac{1}{2}\\
\frac{1}{2}&\frac{1}{2}&-1&\frac{1}{2}&\frac{1}{2}
\end{pmatrix}
=
\begin{pmatrix}
\tilde{S}_{11}&\tilde{S}_{14}&\tilde{S}_{12}+\tilde{S}_{13}&\tilde{S}_{14}&\tilde{S}_{11}\\
\tilde{S}_{21}&\tilde{S}_{24}&\tilde{S}_{22}+\tilde{S}_{23}&\tilde{S}_{24}&\tilde{S}_{21}\\
\tilde{S}_{31}&\tilde{S}_{34}&\tilde{S}_{32}+\tilde{S}_{33}&\tilde{S}_{34}&\tilde{S}_{31}\\
\tilde{S}_{41}&\tilde{S}_{44}&\tilde{S}_{42}+\tilde{S}_{43}&\tilde{S}_{44}&\tilde{S}_{41}\\
\end{pmatrix}.
\label{eq:S Stilde for double layer ising}
\end{equation}
Note that all $|\tilde{S}_{ab}|=1/2$ since the theory contains only Abelian anyons. Thus, 
\eqnref{eq:S Stilde for double layer ising} determines all matrix elements of $\tilde{S}$, except for
$\tilde{S}_{22}=-\tilde{S}_{23}=-\tilde{S}_{32}=\tilde{S}_{33}$. (The equality $-\tilde{S}_{23}=-\tilde{S}_{32}$ follows from the fact that a modular $S$ matrix is symmetric.)
At the same time, we have from $T n=n \tilde{T}$ that 
$\theta_1=1$, 
$\theta_2=\theta_3=e^{\mathrm{i}\pi/4}$,
$\theta_4=-1$. Furthermore, the (2,2) component of the equation $\tilde{S}^2=\Theta(\tilde{S}\tilde{T})^3$ reads
\begin{equation}
\frac{1}{2}\left(1+4\tilde{S}_{22}^2\right)=\frac{1}{2}\left(1+8\mathrm{i}\tilde{S}_{22}^3\right)
\end{equation}
yielding the unique solution $\tilde{S}_{22}=-\mathrm{i}/2$, that also satisfies $|\tilde{S}_{22}|=1/2$.
We can use the thus obtained $\tilde{S}$ matrix to compute the fusion coefficients from \eqnref{eq: new fusion coefficients}, and we find that they are all non-negative integers. The new fusion rules are
\begin{equation}
2\times2=3\times3=4,\qquad2\times3=1,
\end{equation} 
which are distinct from the toric code fusion rules. The resulting TQFT coincides with the gauged Chern number 2 superconductor from Kitaev's 16-fold way \cite{Kitaev06}. We have thus shown that this TQFT is obtained in a unique way through condensation in a double layer of Ising theories (two gauged Chern number 1 superconductors). In fact, one can iterate this procedure to obtain all TQFTs appearing in Kitaev's 16-fold way. A natural open question is to find out which TQFTs exhibit such a closed structure with unique condensations. Using the formalism developed above, we will show below that another simple non-Abelian TQFT, the Fibonacci category, does not admit a similar structure, since it does not allow for any condensation. 

One may wonder whether \eqnref{eq: new fusion coefficients} needs to be imposed as a separate condition on the possible solutions for $\tilde{S}$, or whether it follows from the other conditions in the above list. To show that \eqnref{eq: new fusion coefficients} is required, we discuss the example of four layers of Ising TQFTs in Appendix~\ref{sec: 4 layers Ising}, for which there exist a unitary and symmetric $\tilde{S}$ matrix, except that the fusion coefficients generated from $\tilde{S}$ by Verlinde's formula in \eqnref{eq: new fusion coefficients} are not integer. Therefore, it does not correspond to an allowed condensation transition and the list of conditions is not complete without \eqnref{eq: new fusion coefficients}. 

\section{Layer constructions and uncondensable bosons}
\label{sec: layer construction}

In this section, we apply the condensation formalism to TQFTs $\mA^{(N)}$ that are tensor products of $N$ identical layers of a TQFT $\mA$. There are several motivations to study such a construction:

(1) Some TQFTs are characterized by a $\mathbb{Z}_m$ grading under layering:
 $N=m$ layers can be physically equivalent to the trivial TQFT in the bulk. 
For a theory to  be condensable to  nothing, $m$ is constrained by the fact that the chiral central charge, which is conserved under condensation, must vanish (mod~8). Condensation provides a way to determine the grading $m$ as well as all the TQFTs for $N=1,\ldots, m$ layers. See the following Sec.~\ref{subsec.ZmGraded} for discussions and details of examples.

(2) The grading of TQFTs has an immediate physical implication: Kitaev's 16-fold way, which we discuss below, characterizes 16 different chiral superconductors in (2+1) dimensions.

(3) Layer constructions have been proposed to gain insight into (3+1)-dimensional phases with topological order, for which there is currently no systematic understanding\cite{JianQi14}. The idea is to couple $N$ layers of a TQFT $\mA$ by a condensation transition in such a way that the number of anyons after condensation does not scale with $N$. Some of the anyons that restrict to deconfined particles have a nontrivial particle in every layer. Their restriction is then interpreted as a string excitation of the (3+1)-dimensional theory. We discuss an example in the following Sec.~\ref{subsec.ZmGraded}.

Before condensation, the general structure of $\mA^{(N)}$ is 
\begin{equation}
S^\prime_{\mA^{(N)}}= \underbrace{S_{\mA}\otimes\cdots\otimes S_{\mA}}_{N\ \text{times}},
\quad
T^\prime_{\mA^{(N)}}= \underbrace{T_{\mA}\otimes\cdots\otimes T_{\mA}}_{N\ \text{times}},
\end{equation}
for the modular matrices and 
\begin{equation}
N^{\prime \bs{c}}_{\bs{a},\bs{b}}=\prod_{i=1}^N N^{c_i}_{a_i,b_i},
\qquad
d^\prime_{\bs{a}}=\prod_{i=1}^N d_{a_i},
\qquad
\theta^\prime_{\bs{a}}=\prod_{i=1}^N \theta_{a_i},
\end{equation}
for the fusion matrices, quantum dimensions, and topological spins. Here, $N^{c}_{a,b}$, $d_a$ and $\theta_a$, are the fusion coefficients, quantum dimensions, and topological spins of  $\mA$ and the respective primed quantities belong to $\mA^{(N)}$. We have labeled the anyons in $\mA^{(N)}$ by a vector
$\bs{a}=(a_1,\cdots, a_N)^{\mathsf{T}}$ of anyons in each layer, $1,\ldots,N$, where each entry $a_i$ can be any of the anyons in $\mA$.

\subsection{Theories with $\mathbb{Z}_m$-graded condensations}
\label{subsec.ZmGraded}

\subsubsection{$\mathrm{SU}(3)_1$: 4-fold way}
As a simple example, let us consider the $\mathrm{SU}(3)_1$ TQFT. It has three Abelian anyons 1, 3, $\bar{3}$ with fusion rules
\begin{equation}
3\times 3=\bar{3},\qquad \bar{3}\times\bar{3}=3,\qquad 3\times\bar{3}=1
\end{equation}
and topological spins $\theta_3=\theta_{\bar{3}}=e^{\mathrm{i}2\pi/3}$.
Now, we consider multiple layers of $\mathrm{SU}(3)_1$. Notice that since each layer has a automorphism symmetry $3\leftrightarrow\bar{3}$, all statements below should be understood modulo this automorphism symmetry applied to every layer.

Clearly, the $m=2$ layer theory $\mathrm{SU}(3)_1\times \mathrm{SU}(3)_1$ has no bosons and therefore no condensation transition is possible. 

The $m=3$ layer theory $\mathrm{SU}(3)_1\times \mathrm{SU}(3)_1\times\mathrm{SU}(3)_1$ has 8 bosons. However, up to the automorphism, there is a unique condensation corresponding to bosons $(1,1,1)$, $(3,3,3)$ and $(\bar{3},\bar{3},\bar{3})$ restricting to the vacuum $1^\prime$ and all other bosons confined. Besides the vacuum, two more particles are deconfined: $3^\prime$ with lifts $(3,\bar{3},1)$, $(1,3,\bar{3})$, $(\bar{3},1,3)$, and $\bar{3}^\prime$ with lifts $(\bar{3},3,1)$, $(1,\bar{3},3)$, $(3,1,\bar{3})$. Together, $1^\prime$, $3^\prime$, and $\bar{3}^\prime$ furnish $\overline{\mathrm{SU}(3)}_1$, which differs from $\mathrm{SU}(3)_1$ by complex conjugation of the topological spins. 
It might seem unusual that the condensation of multiple layers of chiral theories results in an anti-chiral theory, but we remind the reader that the chiral central charge is only conserved modulo 8 under condensation transitions and hence $-2-2-2=2~\text{mod}~8$ is allowed.

Then, the $m=4$ layer theory is $\overline{\mathrm{SU}(3)}_1\times\mathrm{SU}(3)_1$ which can be condensed to the trivial TQFT by condensing simultaneously $(3^\prime,3)$ and $(\bar{3}^\prime,\bar{3})$, which confines all other particles. We have thus shown that condensation induces in a unique way a $\mathbb{Z}_4$ grading in the layered $\mathrm{SU}(3)_1$ TQFTs. 

\subsubsection{Ising: Kitaev's 16-fold way}
\label{subsubsec: Ising}

We want to couple $N$ layers of the Ising TQFT, which is defined in \eqnref{eq: Ising}.
For condensation, the simplest boson that we can build consists of the $\psi$-particles in two consecutive layers $n+1$ and $n+2$,
\begin{equation}
B_n:=(1_n,\psi,\psi,1_{N-n-2}),
\end{equation}
where $1_n$ stands for the vacuum particle in $n$ consecutive layers. All bosons $B_n$, $n=0,...,N-2$, are condensed.
We will identify all bosons of this form with the vacuum, building a simple current condensate. From
\begin{equation}
\begin{split}
&(1_n,\psi,\psi,1_{N-n-2})\times(\cdots,\psi,1,\cdots)	=(\cdots,1,\psi,\cdots ),
\end{split}
\end{equation} 
we see that consistency requires that any pair of anyons
$(\cdots,\psi,1,\cdots)$ and $(\cdots,1,\psi,\cdots)$
restrict to the same anyon after condensation. Here, $\cdots$ stands for any sequence (that agrees between the two particles). 
Furthermore, by fusion with the condensate we have
\begin{equation}\label{eq:ConfinementKitaev16Fold}
\begin{split}
&(1_n,\psi,\psi,1_{N-n-2})\times(\cdots,\sigma,1,\cdots)=(\cdots,\sigma,\psi,\cdots).
\end{split}
\end{equation} 
However, $\theta_{(\cdots\sigma1\cdots)}=-\theta_{(\cdots\sigma\psi\cdots)}$, implying that the restrictions of $(\cdots\sigma,1\cdots)$ are confined, because they 
have another lift $(\cdots\sigma,\psi\cdots)$ with different topological spin. By that argument we have shown that the set $\mathcal{Q}$ consisting of particles with least one and at most $N-1$ $\sigma$'s restricts only to confined particles.  On the other hand, we know that particles containing no $\sigma$'s (i.e., only $1$'s or $\psi$'s) restrict to single deconfined particles: 

\begin{itemize}
\item
By closure of the condensate, any particle with even number of $\psi$ and otherwise 1 restricts to the new vacuum $1^\prime$.
\item
Any particle with odd number of $\psi$ and otherwise 1 restricts to the deconfined particle $\psi^\prime$. Their fusion rule is
\begin{equation}
\psi^\prime\times\psi^\prime=1^\prime.
\end{equation}
\end{itemize}

The only particle left to consider is $\sigma^{(N)}\equiv(\sigma,\ldots,\sigma)$. It is easy to show that $\sigma^{(N)} \times \mathcal{Q} \subseteq \mathcal{Q}$. It then follows from the $a=\sigma^{(N)},b\in \mathcal{Q}$, $t=\varphi$ component of \eqnref{eq.commuteRstrFus} that $\sigma^{(N)}$ and particles in $\mathcal{Q}$ restrict to disjoint sets of particles, because the righthand side of \eqnref{eq.commuteRstrFus} is zero in this case, as none of the particles in $\mathcal{Q}$ restrict to the vacuum. But then the restriction of $\sigma^{(N)}$ cannot possibly contain confined particles as those confined particles would have just a single lift $\sigma^{(N)}$, which is impossible from the definition of confined particle. Hence $\sigma^{(N)}$ restricts only to deconfined particles, and we can identify $\mathcal{Q}$ as the set of lifts of all confined particles. 

We can say more about the restriction of $\sigma^{(N)}$. Note $D_{\mU}=D_{\mA}/q=2^N/2^{N-1}=2$,
because $q$ is equal to the number of condensed bosons i.e., $q=2^{N-1}$. As we already know $1^\prime,\psi^\prime$ are deconfined, $D_{\mU} = \sqrt{1+1+\ldots}=4$, where $\ldots$ are additional contributions from the restriction of $\sigma^{(N)}$. When $N$ is not a multiple of $8$, there are just two options. Either \textbf{case (1)} $(\sigma,\cdots,\sigma)$ splits into just two Abelian particles distinct from $1^\prime,\psi^\prime$, or \textbf{case (2)} $(\sigma,\cdots,\sigma)$ has a single restriction with quantum dimension $\sqrt{2}$.
(When $N$ is a multiple of 8 the $\sigma$-string is itself a fermion or boson and could restrict to the $\psi^\prime$ and the vacuum, respectively. However, by $D_{\mathcal{U}}=2$ it is not possible that $\psi^\prime$ and the $\sigma$-string have a common restriction in the case where $N$ is an odd-integer multiple of 8 (since $D_{\mU}=\sqrt{3}$ in that case). The case where $N$ is a multiple of 16 will be discussed separately below.)
Consider now from \eqnref{eq: nS nT equations} the matrix element that corresponds to any particle $t$ in the restriction of  $(\sigma,\cdots,\sigma)$ and the identity in $\mA$,
\begin{equation}
n^t_{(\sigma,\cdots,\sigma)}=\frac{\sqrt{2}^N}{2}d_t,
\label{eq: restrictions of sigma string}
\end{equation}
since we know from the discussion following \eqnref{eq:ConfinementKitaev16Fold} that $t$ has only one lift, $(\sigma,\cdots,\sigma)$.

From the condition that $n^t_{(\sigma,\cdots,\sigma)}$ is integer, we conclude that \textbf{case (1)} applies to even $N$ and \textbf{case (2)} to odd $N$. We now analyze the two cases separately.

\paragraph{Case: $N$ odd}
According to \eqnref{eq: restrictions of sigma string}, we have 
$(\sigma,\cdots,\sigma)\to 2^{(N-1)/2}\sigma^\prime$. It follows from the fusion rules of the original theory, i.e., from Eq.~\eqref{eq.commuteRstrFus} by choosing $a=b=(\sigma,\cdots,\sigma)$, that
\begin{equation}
\sigma^\prime\times\sigma^\prime=1^\prime+\psi^\prime.
\label{eq: 1string with one restriction}
\end{equation}
Thus, $1^\prime, \sigma^\prime, \psi^\prime$ furnish the same (Ising) fusion algebra as $1,\sigma,\psi$ do in every layer.
The spin factors of the deconfined restrictions are given by
\begin{equation}
\theta_{1^\prime}=1,\qquad \theta_{\sigma^\prime}=e^{2\pi\mathrm{i}\nu/16},\qquad \theta_{\psi^\prime}=-1,
\end{equation}
where $\nu=N\,\mathrm{mod}\,16$ is an odd integer, for $N$ is odd. We have thus obtained all TQFTs with Ising fusion rules that appear in Kitaev's 16-fold way.

\paragraph{Case: $N$ even}
If $N$ is even, \eqnref{eq: restrictions of sigma string}  yields the restriction 
$(\sigma,\cdots,\sigma)\to 2^{N/2-1}a^\prime+2^{N/2-1}b^\prime$ with equal coefficients. 
To find the fusion rules for $a^\prime$ and $b^\prime$, we solve \eqnref{eq: nS nT equations}.
This leaves two possibilities
\begin{eqnarray}
&a^\prime\times a^\prime=b^\prime\times b^\prime=1^\prime,
\qquad
a^\prime\times b^\prime=\psi^\prime,
\label{eq: toric code with a b}
\\
&a^\prime\times a^\prime=b^\prime\times b^\prime=\psi^\prime,
\qquad
a^\prime\times b^\prime=1^\prime.
\label{eq: alternative with a b}
\end{eqnarray}
Here, \eqnref{eq: toric code with a b} are the toric code fusion rules. Which of the two cases applies can be determined from the equation
 $\tilde{S}^2=\Theta (\tilde{S}\tilde{T})^3=\tilde{C}$, by using the topological spins
 \begin{equation}
\theta_{a^\prime}=\theta_{b^\prime}=e^{2\pi\mathrm{i}N/16}.
\end{equation}
For $N=2~\mathrm{mod}~4$ one finds the solution \eqnref{eq: alternative with a b} and for 
$N=4~\mathrm{mod}~4$ one finds the solution \eqnref{eq: toric code with a b}.

The case where $N$ is a multiple of 16 has to be considered separately. The condensation described here leads to the toric code TQFT in which $a^\prime$ and $b^\prime$ are bosons. We have shown above that the toric code can be condensed to the trivial TQFT by condensing either $a^\prime$ or $b^\prime$ (which were called $e$ and $m$ before). Thus, in the case where the $\sigma$ string is a boson, two condensations are possible: one leads to the toric code and in the other one, in which the $\sigma$ string restricts in part to the vacuum, leads to the trivial TQFT. The toric code is also the TQFT that was proposed to describe a gauged $s$-wave superconductor without topological edge modes.~\cite{Sondhi}

Together, this $\mathbb{Z}_{16}$ grading represents Kitaev's 16-fold way,
yielding a (non-)Abelian fusion category for the vortices of even (odd) layer length. 
From the point of view of layer construction\cite{JianQi14}, we note that $\psi^\prime$ is a point-like fermionic excitation in 3D space, while $\sigma^\prime$, $a^\prime$ and $b^\prime$ are to be interpreted as vortex or line-like excitations in 3D, because their lift has a nontrivial anyon in each layer. 

It is tempting to consider the topological orders that have been proposed in Refs.~\onlinecite{Fidkowski1,Fidkowski2} as the possible symmetry-preserving gapped surface terminations of time-reversal symmetric (3+1)-dimensional superconductors as another example of a theory with $\mathbb{Z}_{16}$ grading under condensation. The topological index $\nu$ of the bulk superconductor has been shown to be only meaningful mod 16 in the presence of interactions. The $\nu=1$ surface topological order was proposed to be nonmodular category SO(3)$_6$, while that for $\nu=2$ is the so-called T-Pfaffian state. We do not further elaborate on possible condensations in this theory here, as the focus of the present work is on condensation in modular categories.  However, if we were to apply the formalism of \eqnref{eq: nS nT equations} to this problem, none of the possible condensation transitions in a double layer SO(3)$_6\times$SO(3)$_6$ would lead to the T-Pfaffian. Rather, one can condense all bosons in SO(3)$_6\times$SO(3)$_6$ to obtain the trivial nonmodular TQFT $\{1,f\}$ with only one Abelian fermion $f$.

\subsection{Theories with $Z$-fold way: Fibonacci TQFT}
\label{sec: Fib}

Not every TQFT has a $\mathbb{Z}_m$-graded structure under condensation. The simplest counter-example is the Fibonacci TQFT with the single nontrivial anyon $\tau$ and the fusion rule
\begin{equation}
\tau\times\tau=1+\tau.
\end{equation} 
It has topological spin $\theta_\tau=e^{\mathrm{i}4\pi/5}$ and quantum dimension $d_\tau=\phi$, where $\phi=(1+\sqrt{5})/2$ is the golden ratio. 

First, we want to show that no condensation is possible in 5 layers of Fibonacci, despite the presence of the boson $(\tau\tau\tau\tau\tau)$. We will show that there is no matrix $M$ that describes a condensation and satisfies \eqnref{eq: M, T commutator}. To see this, consider the ($1$,b) component of the equation $MS_{\mathrm{Fib}^{(5)}}=S_{\mathrm{Fib}^{(5)}}M$, 
\begin{equation}
\sum_an^\varphi_a\left(S_{\mathrm{Fib}^{(5)}}\right)_{a,b}=\frac{1}{(2+\phi)^{5/2}}\sum_ad_aM_{a,b}.
\end{equation} 
Observe that the righthand side is nonnegative for any $b$. Specializing to $b=(\tau,1,1,1,1)$, we find the lefthand side
\begin{equation}
(2+\phi)^{-5/2}\left(\phi-n^\varphi_{(\tau\tau\tau\tau\tau)}\phi^4\right),
\end{equation} 
which is negative for any $n^\varphi_{(\tau\tau\tau\tau\tau)}\geq 1$, i.e., for any condensation.
Therefore, no condensation transition is possible in 5 layers of Fibonacci (see Ref. \onlinecite{Bais2} for an alternative proof). 

Second, let us show further that no condensation is possible in 10 layers of Fibonacci. Besides the vacuum, there is a  boson with a $\tau$ anyon in every layer, which we denote by $(10\tau)$, and $252={10 \choose 5}$ bosons with $\tau$ anyons in exactly 5 layers. Again, we will show that there is no matrix $M$ that describes a condensation and satisfies \eqnref{eq: M, T commutator}. To see this, we consider the ($1$,b) component of the equation $MS_{\mathrm{Fib}^{(10)}}=S_{\mathrm{Fib}^{(10)}}M$, but this time for the choice $b=(10\tau)$. Up to an overall factor of the total quantum dimension, the equation reads
\begin{equation}
\begin{split}
&n^\varphi_1\phi^{10}+
\sum_{a\in5\tau \ \text{bosons}}(-1)^5\phi^5 n^\varphi_a 
+(-1)^{10}n^\varphi_{(10\tau)}\\
&\quad=n^\varphi_{(10\tau)}+
\sum_{a\in5\tau \ \text{bosons}}\phi^5M_{a,(10\tau)} +
\phi^{10}M_{(10\tau),(10\tau)}.
\end{split}
\end{equation}
Using $n^\varphi_1=1$, it simplifies to
\begin{equation}
\begin{split}
0=
\phi^5\left(
M_{(10\tau),(10\tau)}-1
\right)
+
\sum_{a\in5\tau \ \text{bosons}}
\left(
n^\varphi_a+M_{a,(10\tau)}
\right)
.
\end{split}
\label{eq: 10 layer equation}
\end{equation}
We can see that Eq.~\eqref{eq: 10 layer equation} has no nontrivial solution: Since $\phi^5$ is irrational, the first term needs to be zero on its own, which requires $M_{(10\tau),(10\tau)}=1$. This implies that $(10\tau)$ does not condense, as it has noninteger quantum dimension and would therefore have to split in order to condense. However, the second term in Eq.~\eqref{eq: 10 layer equation} is a sum of nonnegative numbers that can only vanish if $n^\varphi_a=0, \ \forall a$. Hence, none of the bosons condenses.

In fact, one can show that no condensation is possible for \emph{any} number of layers $N$  of the Fibonacci TQFT~\cite{Bookera12}. We will reformulate this proof much more easily using the formalism developed in this paper elsewhere in a way that also generalizes to other TQFTs~\cite{He15}.

\section{Conclusions}

In summary, we derived a framework for the condensation of anyons that is applicable to 
 modular tensor category models of topological order. Our derivation is based on a small number of physical assumptions and focuses on the computation of the modular matrices $\tilde{S}$ and $\tilde{T}$ of the theory after condensation. Based on this, we propose an algorithm to carry out this computation. This algorithm first seeks symmetric nonnegative integer matrices $M$ that commute with the modular matrices $S$ and $T$ of the original theory. It  then proceeds by factorizing $M=n n^{\mathsf{T}}$ in a product of a nonnegative integer matrix $n$ with itself. Finally, the equations $S n= n \tilde{S}$ and $T n=n \tilde{T}$ are solved. Our algorithm has proven to be practically useful in all examples that we studied. We finally demonstrated that the equations that are central to our derivation are powerful constraints on condensation transitions in general.

This leads us to several open problems that are not answered by the present work. One concerns the assumption that $\beta_t=0$ for all confined particles $t$. We have shown in Secs.~\ref{sec.simplecurrent} and~\ref{sec: One confined particle theories} that this relation follows from weaker assumptions for certain theories. But a general proof of this statement is lacking, so that it remains an assumption for us.
Other questions concern the uniqueness of solutions and the transitivity of condensation transitions. For example, given an $M$, is there a unique $n$ that solves $M=n n^{\mathsf{T}}$ and leads to a valid condensed theory? And given such a solution $n$, is there a unique consistent solution $\tilde{S}$ and $\tilde{T}$? In a similar vein, is the condensed theory completely characterized by the coefficients $n^{\varphi}_{a}$?\footnote{Indeed, we cannot exclude the possibility that  additional information, like certain vertex lifting coefficients, are needed to fully determine the topological order of the condensed phase.} At present, we do not have counterexamples against affirmative answers to these questions.

Another future direction could be the condensations in the presence of global symmetries\cite{BBCW14}. When we have global symmetries on top of a topologically ordered system, the anyons may transform in a projective representation. A direct consequence is that certain condensations may not be able to happen if all global symmetries are respected.

\section{Acknowledgement}

The authors thank Lukasz Fidkowski, Yidun Wan, Parsa Bonderson, Roger Mong and Jeffrey Teo for discussions. TN and CVK are supported by the Princeton Center for Theoretical Science. GS acknowledges support of  Fulbright grant PRX14/00352,  MINECO grant FIS2012- 33642, CAM grant  QUITEMAD+ S2013/ICE-2801,  and the 
grant SEV-2012-0249 of the  ``Centro de Excelencia Severo Ochoa" Programme. BAB acknowledges support of MURI-130-6082, ONR-N00014-11-1-0635, NSF CAREER
DMR-0952428, NSF-MRSEC DMR-0819860, the Packard Foundation, and a Keck grant.

\section{Appendices}

\appendix

\section{Essentials of modular tensor categories}
\label{app.MTC}

In this appendix, we present a short review of the modular tensor category description of a (2+1)-dimensional TQFT. This approach only describes the low energy excitations of the TQFT, i.e., the anyons. The anyons are usually labeled by objects $a,~b,~c, \ldots$ and are supplemented by other data, such as the fusion coefficients $N_{ab}^c$. For a comprehensive overview of the category theory approach of TQFT, we refer the reader to Refs.~\onlinecite{Kitaev,Bonderson07,BernevigNeupert15}. Here, we only present a brief and simple review of the important properties that we frequently use in this paper.

\subsection{Fusion rules and quantum dimension}

The anyons of a TQFT can fuse. When two anyons come close to each other spatially, they can fuse into other anyons. An analogy can be drawn to the algebra of spins: if we take two spin $\textstyle{\frac{1}{2}}$ particles, they can fuse into either spin $0$ and spin $1$ particle. For this, we would write, in group representation theory, the fusion rule $\textstyle{\frac{1}{2}\times\frac{1}{2}=0 + 1}$. In general, the fusion of anyons in a TQFT is represented via
\begin{eqnarray}
a \times b = \sum_{c} N_{ab}^{c} c,
\end{eqnarray}
where $a,~b,~c$ are labels for the anyons, and the fusion coefficients $N_{ab}^c$ are non-negative integers. 
The fusion can be represented by a state $|a,b;c,\mu\rangle$ in the fusion vector space $V^{ab}_c$. Here $\mu=1,\cdots, N^c_{ab}$ labels the vectors that form a basis of the $N^c_{ab}$-dimensional fusion vector space $V^{ab}_c$.

Just like the fusion of spins, we require that the fusion rules are symmetric or commutative, that is, $a\times b$ is equivalent to $b\times a$. This translates to
\begin{eqnarray}
N_{ab}^c=N_{ba}^c.
\end{eqnarray}

Moreover, fusion rules are also associative. Suppose we take three anyons $a,~b,~c$ and try to fuse them. Then we have two ways to do so: $(a\times b)\times c$ and $a\times (b\times c)$. We require the fusion rule to be associative by requiring that the two fusions yield the same result. In terms of the fusion coefficients, this translates to 
\begin{eqnarray}
\sum_{d,e} N_{ab}^d N_{dc}^e=\sum_{d,e} N_{ad}^e N_{bc}^d.
\label{eq: N commute}
\end{eqnarray}

Other important data associated with anyons are their so-called quantum dimensions $d_a,~d_b, \cdots$. This concept appears because anyons are associated with nontrivial internal Hilbert spaces. Again, we can take the example of spins to illustrate this. In the case of spin $\textstyle{\frac{1}{2}}$, where $\textstyle{\frac{1}{2} \times \frac{1}{2} = 0 + 1}$, spin $\textstyle{\frac{1}{2}}$ is associated with a two-dimensional Hilbert space, and meanwhile spin $0$ is associated with a one-dimensional Hilbert space, spin $1$ a three-dimensional Hilbert space. As we can see, the total dimension of Hilbert space does not change after fusion. The product $\textstyle{\frac{1}{2} \times \frac{1}{2}}$ has a $(2\times 2=4)$-dimensional Hilbert space while $\textstyle{0+1}$ has a $(1+3=4)$-dimensional Hilbert space. Similarly, in a TQFT, we also have
\begin{eqnarray}\label{eq.quantdim}
d_a d_b = \sum_{c} N_{ab}^c d_c.
\end{eqnarray}
The above equation can be viewed as an eigenvalue equation of a matrix $N_a$ whose entries are $(N_a)_{bc}=N_{ab}^c$. The eigenvector is $(d_b)$, the eigenvalue is $d_a$. Equation~\eqref{eq: N commute} says that all matrices $N_a, N_b,\ldots$ commute and thus they have common eigenvectors, one of which is the vector of all quantum dimensions. The total quantum dimension of a TQFT $D$ is defined as the norm of the quantum dimension vector, $D=\sqrt{\sum_a d_a^2}$.

If all anyons of a TQFT have quantum dimension 1, we call such a theory Abelian. If there exist anyons with quantum dimension larger than 1, we call such a theory non-Abelian. This is intimately related to the Perron-Frobenius theorem, where $d_a$ is a Frobenius eigenvalue, and hence has to satisfy $\min_b \sum_c (N_a)_{bc}\leq d_a \leq \max_b \sum_c (N_a)_{bc}$. Hence $d_a>1$ implies that there exists a $b$ such that $ \sum_c (N_a)_{bc}>1$, so $a\times b$ contains more than one particle.

\subsection{Braiding, topological spin and modular matrices}

Another physically important concept in a TQFT is braiding. This allows us to determine how a state transforms when its anyons are adiabatically moved around each other. In Abelian theories, when we adiabatically move an anyon $a$ fully around another anyon $b$, the state transforms through multiplication by a universal monodromy phase. For example, if we take a fermion around a $\pi$ flux, the wave function obtains a topological minus sign $-1$. Another special case is when we exchange two identical abelian anyons $a$. This process defines the topological spin $\theta_a$ for the particle $a$.

In non-Abelian theories, the braiding operation $\mathcal{R}_{ab}$ between two anyons $a$ and $b$ is an operator that acts on the Hilbert space $V^{ab}_c$ which describes states of $a$ and $b$ that fuse into a fixed anyon $c$. If we denote a basis of $V^{ab}_c$ by $|a,b;c,\mu\rangle$ with $\mu=1,\cdots, N^{c}_{ab}$, then $\mathcal{R}_{ab}$ has the representation
\begin{eqnarray}
\mathcal{R}_{ab}|a,b;c,\mu\rangle=\sum_{\nu} [R^{ab}_c]_{\mu\nu} |b,a;c,\nu\rangle.
\end{eqnarray}
In this notation, the topological spin for an anyon $a$ is defined as
\begin{eqnarray}
\theta_a=\frac{1}{d_a} \sum_c d_c \mathrm{Tr}_c[R^{aa}_c],
\end{eqnarray}
where $\mathrm{Tr}_c[\cdots]$ is the trace taken in the fusion vector space $V_c^{a a}$.

\begin{subequations}
Given the  braiding $\mathcal{R}_{ab}$, we can construct the modular matrices $S$ and $T$ which are the same modular matrices encoding the global data of $S$ and $T$ in a CFT. They are given by
\begin{eqnarray}
S_{ab}&=& \sum_{c} N_{ab}^c \mathrm{Tr}[R^{ab}_c R^{ba}_c] d_c,	\\
T_{ab}&=& \theta_a \delta_{ab}.
\end{eqnarray}
\end{subequations}
By definition, $S$ is a symmetric matrix. Moreover, in a modular tensor categories, $S$ and $T$ are unitary matrices satisfying $S^\dag S=S S^\dag=1$, $T^\dag T=T T^\dag=1$.

In Refs.~\onlinecite{Bais3,HHW,MW15,LWYW13}, the $S$ matrix is used as an order parameter to detect topological phase transitions and anyon condensations. The implicit assumption in doing so is that the $S$ matrix represents physical, measurable properties of the state, unlike, say, the gauge-dependent $F$-symbol, which is another MTC data that we will not introduce here.

\section{Quantum dimensions of $\mA$ and $\mT$}
\label{app.QutmDim}

\subsection{Proof of $d_a=\sum_{r\in\mT} n^r_a d_r$}

From the \eqnref{eq.commuteRstrFus} we obtain, by multiplying both sides by the quantum dimension $d_t$ of the particle $t$ in the $\mT$ theory and summing over $t$
\begin{equation}
\begin{split}
\sum_{r,s,t\in\mT} n_a^r n_b^s \tilde{N}_{rs}^t d_t =& \sum_{c\in\mA,t\in\mT} N_{ab}^c n_c^t d_t \\
=& \sum_{r,s\in\mA} n_a^r n_b^s d_r d_s,
\end{split}
\end{equation}
where we are only considering $\mT$ theories which are also fusion categories (not braided ones) and hence satisfy the equivalent \eqnref{eq.quantdim} for the $\mT$ theories
\begin{eqnarray}
\sum_{t\in\mT} \tilde{N}_{rs}^t d_t = d_r d_s.
\end{eqnarray} 

We then have the trivial re-writing
\begin{eqnarray}
\sum_{c\in\mA} N_{ab}^c \left( \sum_{t\in\mT} n_c^t d_t  \right)  = \left( \sum_{r\in\mT} n_a^r  d_r \right) \left(  \sum_{s\in\mT} n_b^s d_s \right),
\end{eqnarray} 
which means that $\left( \sum_{t\in\mT} n_c^t d_t  \right)$ is an eigenvalue of the matrix $(N_a)_{bc}$ with eigenvalues and eigenvector $\left( \sum_{r\in\mT} n_a^r  d_r \right)$ and $\left(  \sum_{s\in\mT} n_b^s d_s \right)$, respectively. Since the eigenvector has positive entries, by the Perron-Frobenius theorem, the eigenvalue is the largest eigenvalue of the $N_a$ matrix, and hence it is indeed $d_a$
\begin{eqnarray}
d_a =\sum_{r\in\mT} n_a^r d_r. \label{quantumdimrestriction}
\end{eqnarray}

\subsection{Proof of $d_r=\frac{1}{q}\sum_{a\in\mA} n^r_a d_a$}

We start with Eq.~\eqref{eq.commuteRstrFus}, multiply by $d_a$ and sum over $a\in\mA$. Using Eq.~\eqref{eq.quantdim}, it follows
\begin{equation}
\begin{split}
\sum_{r,s\in\mT}\sum_{a \in\mA} n_a^r d_a \tilde{N}_{rs}^t n_b^s 
=&\,\sum_{a,c\in\mA} n_c^t N_{ab}^c d_a	\\
=&\, d_b \sum_{c\in\mA} n_c^t d_c.
\end{split}
\end{equation} 

For the simplicity of notations, let $\alpha_t\equiv\sum_{c\in\mA} d_c n_c^t$, which satisfies the eigenvalue equation
\begin{eqnarray}\label{eq:b6}
\sum_{r\in\mT} \left(\sum_{s\in\mT} n_b^s \tilde{N}_s\right)_{tr} \alpha_r = d_b \alpha_t.
\end{eqnarray} 
Notice the unorthodox use of the matrix $(\tilde{N}_s)_{tr} = \tilde{N}_{sr}^t$, unlike in the line following Eq.~\eqref{eq.quantdim}. We define the matrix this way in order not to use the equation $n_a^t = n_{\bar{a}}^{\bar{t}}$. This matrix has the vector of quantum dimensions $(d_1,\ldots, d_{N})^{\mathsf{T}}$, where $N$ are the number of particles in the fusion category $\mT$, as an eigenvector $\forall s$ in $(\tilde{N}_s)_{tr}$. Since we are using the $\tilde{N}_s$ matrix in an unorthodox fashion (it is the transpose of the usual $\tilde{N}_s$ matrix), we prove the statement
\begin{eqnarray}
\sum_t \tilde{N}^r_{st} d_r = d_s d_t = \sum_t \tilde{N}^{\bar{t}}_{s\bar{r}} d_r = d_s d_{\bar{t}}.
\end{eqnarray} 
It follows from the above that $\sum_r (\tilde{N}_{s})_{tr} d_r = d_s d_t$. Hence  $(d_1,\ldots, d_{N})^{\mathsf{T}}$ is a common eigenvector of all the $\tilde{N}_s$, even as defined in the unusual way above.  

We now sum Eq.~\eqref{eq:b6} over $b$ to get
\begin{eqnarray}
\sum_{r\in\mT} \left(\sum_{b\in\mA}\sum_{s\in\mT} n_b^s \tilde{N}_s\right)_{tr}  \alpha_r 
=\left(\sum_{b\in\mA} d_b\right) \alpha_t.
\end{eqnarray} 
The matrix $(\sum_{b\in\mA}\sum_{s\in\mT} n_b^s \tilde{N}_s)_{tr}$ is a completely positive matrix with integer strictly positive coefficients: for any $t,r$, there exists $s$ such that $\tilde{N}_{sr}^t>0$ and for every $s$ there exists an $n_b^s>0$. As such, it satisfies a stronger version of the Perron-Frobenius theorem which says that there is a unique eigenvector with all elements positive, and all other eigenvectors have at least one negative element. As such, since $\alpha_t$ is all positive, we identify it as the unique largest eigenvector. But since the $\tilde{N}_s$ have a common eigenvector, the quantum dimensions of the condensed theory, we then can identify this eigenvector with
\begin{eqnarray}\label{eq:corrofFS}
\alpha_t = \sum_{c\in\mA} d_c n_c^t = q d_t,
\end{eqnarray} 
where $q$ is a proportionality constant. We now find two expressions for it. First, multiplying \eqnref{eq:corrofFS} by $d_t$ and summing over $t$ gives
\begin{equation}
 q\sum_{t\in\mT} d_t^2 = \sum_{c\in\mA} d_c \sum_{t\in\mT} n_c^t d_t = \sum_{c\in\mA} d_c^2,
\end{equation} 
where the last equality follows from \eqnref{quantumdimrestriction}. This implies
\begin{equation}
q= D_\mA^2/D_\mT^2.
\end{equation}
Furthermore, multiplying Eq.~\eqref{eq.commuteRstrFus} by $d_a d_b$ for $t=\varphi$ and summing over $a,b$ reads 
\begin{equation}
\sum_{c\in\mA} N_{ab}^c n_{c}^\varphi = \sum_{t\in\mT} n_a^t n_b^{\bar{t}},
\end{equation}
which implies
\begin{equation}
\begin{split}
\sum_{a,b,c\in\mA} d_a d_b N_{ab}^c n_{c}^\varphi	
= &\, \sum_{t\in\mT} \sum_{a\in\mA} d_a  n_a^t \sum_{b\in\mA} d_b n_b^{\bar{t}} 	\\
=&\, q^2 D_\mT^2.
\end{split}
\end{equation}

On the other hand,
\begin{eqnarray}
\begin{split}
q^2 D_\mT^2 &= \sum_{a,b,c\in\mA} d_a d_b N_{ab}^c n_c^\varphi	\\
&= \sum_{b,c\in\mA} d_b^2 d_c n_c^\varphi	\\
&= D_\mA^2 \sum_{c\in\mA} d_c n_c^\varphi,
\end{split}
\end{eqnarray}
hence 
\begin{eqnarray}
q= \sum_{c\in\mA} d_c n_c^\varphi.
\end{eqnarray}

\section{Chiral algebra}
\label{sec.chiralextension}

In this section, we review the connection between the above formalism and CFT. As pointed out by Bais and Slingerland \cite{Bais1}, the mathematics of boson condensation has a parallel in conformal field theories. First, for at least some MTCs $\mA$, the particle labels are in one-to-one correspondence with the conformal families in some (not necessarily unique) CFT. (The MTC-conformal block correspondence generalizes Witten's  work \cite{WittenJonesPoly} on the relationship between Chern-Simons theory and chiral Wess-Zumino-Witten models.) Second, when this correspondence holds, the process of condensation in the TQFT is closely related to extending the chiral algebra in the CFT \cite{yellow}. 

Let us consider a CFT with a chiral algebra $A$ which contains the stress-tensor $T(z)$ and all locally commuting holomorphic operators
in the theory such as currents $J^a(z)$ associated to Lie groups, etc. The mode expansions of these operators give rise to 
infinite dimensional algebras, like  Virasoro, Kac-Moody or $W$-algebras. The  irreducible representation spaces of the chiral  
algebra $A$,  denoted by ${\cal H}_a$,  are labelled by the primary fields $a$, whose number is finite in a RCFT.
The primary fields are in one-to-one correspondence with the anyons  of a TQFT. The TQFT is nothing but the CFT reduced to its  basic topological data like braiding and fusion matrices, etc. (However, due to this reduction, several distinct CFTs may correspond to the same TQFT.)

For  each  representation space ${\cal H}_a$ there is  a  character 
\begin{eqnarray}
\chi_a(\tau) = {\rm Tr}_{{\cal H}_a} e^{ 2 \pi i \tau(  L_0 - c/24)},
\label{cft1}
\end{eqnarray} 
given by the partition function of the states in ${\cal H}_a$ propagating along a torus with  modular parameter $\tau$ (with ${\rm Im} \, \tau >0)$. The constant $c$ is the central charge of the CFT and $L_0$ is the zero element of the Virasoro algebra. The modular transformations act on the characters as 
\begin{eqnarray}
\chi_a(\tau+1)  & = &  \theta_a  e^{ - \frac{ i \pi c}{12}} \,  \chi_a(\tau), \label{cft2} \\ 
\chi_a \left(- \frac{1}{ \tau} \right)  &  = &  \sum_{b} S_{a b} \chi_b(\tau)  , 
\nonumber 
\end{eqnarray} 
where $\theta_a = e^{ 2 \pi i  h_a }$ is the topological spin, $h_a$ the conformal weight of the primary field $a$.  
The full CFT also contains an anti-chiral algebra, $\bar{A}$, which 
for simplicity we assume to be isomorphic to $A$. Correspondingly,  the  complete Hilbert space is the tensor
product ${\cal H} = \oplus_a {\cal H}_a \otimes \bar{ \cal H}_a$ and the total partition function is
\begin{eqnarray}
Z_{\rm diag} (\tau, \bar{\tau}) = \sum_a  \bar{\chi}_a( \bar{\tau})    \chi_a ( \tau) ,
\label{cft3}
\end{eqnarray} 
which is modular invariant thanks to the $S$, $T$ unitarity: $S S^\dagger = T T^\dagger = \openone$. 
The {\em pairing}  between the left and right states of a non-chiral  CFT can be more general than (\ref{cft3})
\begin{eqnarray}
Z(\tau, \bar{\tau}) = \sum_{a, b}   \bar{\chi}_a( \bar{\tau}) \,  M_{a b}   \,    \chi_b ( \tau),
\label{cft4}
\end{eqnarray} 
where $M$ is called the mass matrix which must satisfy $[S, M] =[ T, M] =0$ to guarantee
the modular invariance of the partition function~(\ref{cft4}).  A fundamental problem in RCFT
is  to classify all possible modular invariant partition functions, that is, mass matrices $M$.
This program has been achieved for theories with  simple currents \cite{GS1,GS2,KS94}, but it is far from being solved in general. 

There are three types of mass matrices: i) Those associated to automorphisms of the fusion rule algebra,  ii) those corresponding to a chiral extension of  $A$, and iii) a combination of i) and ii). This  result is related to  the naturality theorem due to  Moore and Seiberg: {\em In a CFT when the  left and right chiral algebras are maximally extended the field content matrix defines an automorphism $\omega$ of the fusion rule algebra, i.e.: $M_{a, b} = \delta_{a, \omega(b)}$} \cite{MS89b}.
A chiral algebra  $A  \otimes \bar{A}$ is called maximally extended when it includes all the holomorphic and antiholomorphic fields in ${\cal H}$ (i.e., those with integer conformal weights).~\cite{MS89} 

The mass matrices and the associated naturality theorem have a precise correspondence within the boson condensation encountered in the main text. Let us explain it in more detail.

An extension of the chiral algebra  $A$ can arise if there exists a subset $\{ \gamma_ i \}$ of primary fields with integer conformal weights that are mutually local. One can therefore add these holomorphic fields to those already included in $A$ to obtain an extended chiral algebra $U$. 
It is then clear that the representation spaces of the new algebra $U$ should be a combination  of those
of the original algebra $A$. 
In particular, the (irreducible) conformal family vector space ${\cal H}_\varphi$ corresponding to the new identity representation $\varphi$
will be  the direct sum of the {\em old}  identity conformal family ${\cal H}_{1}$ plus  the conformal families corresponding to the old primaries $\gamma_i$, that is ${\cal H}_\varphi =   {\cal H}_{1}  \oplus_i {\cal H}_{\gamma_i}$. The fields $\gamma_i$ correspond  to the bosons that condense in the TQFT.  The space ${\cal H}_\varphi$ is the CFT version of the vacuum after condensation.

The irreducible representation spaces  of the extended chiral algebra $U$,  denoted by ${\cal H}_u$, 
break down into the direct sum of  irreducible representations ${\cal H}_a$  of the smaller algebra $A$. Such
decompositions are called branching rules and are noted as
\beq
{\cal H}_u \rightarrow \oplus_{a \in  \mA} n_a^u {\cal H}_a . 
\label{C5bis}
\eeq
The branching coefficient $n_a^u$ gives the multiplicity of the irreducible representation
$a$  of $ A$  in the decomposition of the irreducible representation $u$ of $ U$. 
The fields appearing in the decomposition (\ref{C5bis}) have to be  mutually local with respect to the
fields in the chiral algebra $U$. From Eqs.~\eqref{C5bis} and.~\eqref{cft1}  follows the expression
for  the  character of the representation $u$  in terms of the characters of the representations $a$ [recall Eq.(C1)]~\cite{yellow} 
\begin{eqnarray}
\tilde{\chi}_u(\tau)  = \sum_{a \in { \mA}} n^u_a  \,  \chi_a ( \tau) , \quad u \in {\mU}. 
\label{cft5}
\end{eqnarray} 
The primary field $u$ corresponds to a deconfined anyon in the TQFT. 
The TQFT Eq.~(\ref{deconfined}) means in CFT that the primary fields that built up a 
representation of the extended algebra must have the same conformal weights  modulo integers. 
On the other hand, if a field $a$ is such that  the orbit $\gamma_i \times a, \; \forall i$
contains fields with different conformal weights, then they disappear  from the representation
theory of $ U$. These fields are associated to the confined anyons  defined in Eq.~(\ref{confined}). 
Given the characters~(\ref{cft5})  of the  extended  chiral algebra ${U}$,  one can construct the
diagonal partition function 
\begin{equation}
\tilde{Z}(\tau, \bar{\tau})  = \sum_{u \in {\mU}}  | \tilde{\chi}_u(\tau) |^2  , 
\label{cft6}
\end{equation}
which when written in terms of the characters~(\ref{cft1}) of ${A}$ reads like Eq.~(\ref{cft4})
with 
\begin{equation}
M_{a b} = \sum_{u \in {\mU}}  n^u_a   \,  n^u_b   . 
\label{cft7} 
\end{equation}
This equation  shows  that an extension of the chiral algebra gives rise to an off-diagonal
partition function and in turn to a boson condensation in the TQFT.  

The original and chirally extended CFTs are both assumed to be modular theories, 
with their characters transforming under modular transformation $S$ and $T$ of the torus parameter $\tau$ as
\begin{equation}
\begin{split}
\tilde{\chi}_s\left(-\frac{1}{\tau}\right)=& \sum_{t} \tilde{S}_{st} \tilde{\chi}_t(\tau)=\sum_{t,a} \tilde{S}_{st} n^{t}_{a}\chi_a(\tau)	\\
=&\sum_b n^{s}_{b}\chi_b\left(-\frac{1}{\tau}\right)=\sum_{a,b} n^{s}_{b} S_{ba}\chi_a(\tau) ,
\end{split}
\end{equation}
i.e.,
\begin{subequations}
\begin{equation}\label{eq.modinvS}
n\tilde{S}=Sn .
\end{equation}
Similarly 
\begin{equation}\label{eq.modinvT}
n\tilde{T}=Tn , 
\end{equation}
where $\tilde{S}$ and $\tilde{T}$ are modular matrices for the ${ U}$ algebra. 
Equation~\eqref{eq.modinvS} and \eqnref{eq.modinvT} also appear as matching conditions in the study of gapped domain walls between two topological phases~\cite{LWW15}.  
\end{subequations}

One can easily deduce that $[M,S]=[M,T]=0$. Moreover, through \eqnref{eq.modinvS} and \eqnref{eq.modinvT}, we can show that
\begin{enumerate}
\item $\tilde{c}=c \text{ (mod 24)}$,
\item $\theta_s=\theta_a$, if $n^s_a\neq 0$,
\item $n \tilde{C}=C n$,
\item $q\equiv\sum_a n^{\varphi}_a d_a=D_{\mathcal{A}}/D_{\mathcal{U}}$,
\item $d_t=\frac{1}{q}\sum_{a\in\mA} n^t_a d_a$,
\end{enumerate}
where $\tilde{C}$ and $C$ are the charge conjugation matrices for the $\mU$ and $\mA$ theories respectively, and $D_{\mathcal{U}}$ and $D_{\mathcal{A}}$ are total quantum dimension of the $\mU$ and $\mA$ theory, respectively.

So far we have discussed the mass matrices that correspond to extensions of the chiral algebra. 
The other possibility is that the mass matrix is a permutation $P$  of the irreducible representations of ${ A}$ that 
 corresponds to an automorphism 
of the fusion rules \cite{yellow}. This case does not describe boson  condensation. The third  possibility is that 
the mass matrix describes an off diagonal partition function of the chiral algebra ${ U}$,  namely $M = n  \tilde{P} n^{\mathsf{T}}$, with $\tilde{P}$ a permutation automorphism of the fusion rules of ${\cal U}$. These possibilities were mentioned before in connection with the Moore and Seiberg naturally theorem  \cite{MS89b}. 

 The conclusions we obtain above, including \eqnref{eq.modinvS} and \eqnref{eq.modinvT}, can be viewed as necessary conditions for boson condensation. So, a solution of the above consistency equations does not guarantee the existence of a boson condensation $\mA \rightarrow \mU$. It could still happen, for example, that the fusion coefficients derived from such a solution via the Verlinde formula are not integers (see Appendix~\ref{sec: 4 layers Ising} for an example). Then, the solution has to be discarded. However, the absence of a solution does imply that there is no boson condensation $\mA \rightarrow \mU$.

\section{Condensations in $SU(2)$ CFTs}
\label{app: SU2 condensation}

To illustrate some properties of the condensation transition we consider the family of CFTs that correspond to SU(2) at level $k$. These theories have $(k+1)$ primary fields in corresponding conformal blocks labelled by integers $a=0,\ldots, k$ that are denoted as $\phi_a$, and the corresponding conformal characters are denoted as $\chi_a$. (In the corresponding TQFT, the
anyon with $a=0$ is the vacuum.) The matrix elements of the modular $S$ and $T$ matrices are given by
\begin{equation}
S_{ab}=\sqrt{\frac{2}{2+k}}\sin\frac{\pi (a+1)(b+1)}{k+2},
\end{equation} 
and
\begin{equation}
T_{ab}= 
e^{2\pi\mathrm{i}\frac{a(a+2)}{4(k+2)}}\delta_{ab}, \quad c = \frac{3 k}{k+2}.
\end{equation}
 All the modular invariant partition functions of this CFT were obtained
by Cappelli, Itzykson and Zuber who found a surprising correspondence with the ADE classification of Lie groups \cite{ADE}. 
The complete list is 
\begin{subequations}
\barray 
 Z_{A_{k+1}} & = &    \sum_{n=0, n \in Z}^k  |\chi_n |^2 \label{ad1} , \\
Z_{D_{2 \ell +2}}   & = &    \sum_{n=0, n \in 2 Z}^{ 2 \ell -2}  |\chi_n  + \chi_{4 \ell -  n}|^2 + 2 | \chi_{2 \ell} |^2 ,  \nonumber   \\
 Z_{D_{2 \ell +1}}  & = &   \sum_{n=0, n \in 2 Z}^{ 4 \ell -2}  |\chi_n|^2   + |\chi_{2 \ell -  1}|^2   \nonumber    \\ 
&  & +  \sum_{n=1, n \in 2 Z +1}^{ 2 \ell- 3} ( \chi_n \bar{\chi}_{4 \ell - 2 - n} + {\chi}_{4 \ell - 2 - n} \bar{\chi}_n)  ,  \nonumber 
\earray
\barray
 Z_{E_{6}}  & = &   |\chi_0  + \chi_{6}|^2 +  |\chi_3  + \chi_{7}|^2 +  |\chi_4  + \chi_{10}|^2 , \nonumber     \\
Z_{E_{7}} &  = &     |\chi_0  + \chi_{16}|^2 +  |\chi_4  + \chi_{12}|^2 +  |\chi_6  + \chi_{10}|^2 
    \nonumber \\ 
& & + |\chi_{8}|^2 + \chi_8 ( \bar{\chi}_2 + \bar{\chi}_{14}) +  ({\chi}_2 + {\chi}_{14}) \bar{\chi}_8 ,  \nonumber    \\
Z_{ E_{8}} &  = &    |\chi_0  + \chi_{10} + \chi_{18}   
+ \chi_{28}|^2 \nonumber \\&&+    |\chi_6  + \chi_{12} + \chi_{16}  + \chi_{22}|^2 \,   , \nonumber 
\earray 
\end{subequations}
where $k=4\ell$ in $Z_{D_{2 \ell +2}}$, $k=4\ell-2$ and in $Z_{D_{2 \ell +1}}$, while $k=10$ in $ Z_{E_{6}}$, $k=16$ in $Z_{E_{7}}$, and $k=28$ in $Z_{ E_{8}}$. Here, $\chi_a$ are the characters of the irreducible representation spaces of the chiral algebra of SU(2)$_k$.

The origin of these  off-diagonal partition functions is the following: 
\begin{itemize}
\item $D_{2 \ell + 2}$: $J=\phi_{4 \ell}$, is a bosonic simple current with integer conformal weight  $h_J = \ell$.  For $\ell=1$, $\phi_4$
is a current  that yields a  chiral extension corresponding  to the conformal embedding $\mathrm{SU}(2)_4 \subset \mathrm{SU}(3)_1$. 
(Notice that the central charge of the two CFTs is the same $c_{\mathrm{SU}(2)_4} = c_{\mathrm{SU}(3)_1}$.) 
\\
\item
 $D_{2 \ell +1}$:  the simple current $J=\phi_{4 \ell}$ has  half-odd  conformal weights, $h_J = \ell- 1/2$, so it 
does not yield an extension of the chiral algebra, i.e., it does not correspond to condensation. The partition function can be written
as $Z_{D_{2 \ell +1}} = \sum_{a} \chi_a \, \bar{\chi}_{\omega(a)}$, where
$\omega$ is the  unique automorphism of the fusion rules, namely $\omega(a) = a$ for $a$ even and $\omega(a) = k-a$ for $a$ odd. 
\\
\item
$E_6$:  chiral extension with the field $\phi_6$ with  $h_6= 1$. This is not a simple current.
The chiral extension corresponds to the  conformal embedding  $\mathrm{SU}(2)_{10} \subset \mathrm{SO}(5)_1$, both CFT's  have the same central charge, namely $c= 5/2$. 
The $\mathrm{SO}(5)_1$ algebra can be constructed with 5 Majorana fermions (i.e. Ising models). In the $\mathrm{SU}(2)_{10}$
theory one has $h_{4} = 1/2, h_{10} = 5/2$, 
 $h_{3} = 5/16,$ $h_{7} = 21/16$, so that $h_{10} - h_4=2$ and  $h_7 - h_3 = 1$. The field $\phi_3$ can be built
from the product of $5$ spin fields  of the Ising model which have $h_\sigma = 1/16$. 
\\
\item
$E_7$: explained by an exceptional automorphism of the $D_{10}$ chiral algebra \cite{MS89b,yellow} [see Eq.(\ref{D10})].   
\\
\item
$E_8$: chiral extension with three operators with $h_{10} = 1,$ $h_{18} = 3,$ and $h_{28} = 7$.  The remaining
fields in $Z_{E_8}$ have weights: $h_{6} = 2/5,$ $h_{12} = 7/5,$ $h_{16} = 12/5,$ $h_{22} = 22/5$. The central charge is $c=  14/5$,
which coincides with that of $G_2$ at level $k=1$ \cite{yellow}.  
\end{itemize}

The results explained above can be summarized in the following table: 
\beq
\begin{tabular}{|c|c|c|c|}
\hline 
Type & $k$ & $Z$ & Comments \\
\hline
$A_{k+1}$ & k & - & - \\
$D_{2 \ell+2}$ &  $ 4 \ell$ & EXT &  $SU(2)_4 \subset SU(3)_1$ \\
$D_{2 \ell+1}$ &  $ 4 \ell -2$ & AUT & -   \\ 
$E_6$ &  10  & EXT  & $SU(2)_{10} \subset SO(5)_1$ \\ 
$E_7$ &  16  & AUT   & -  \\ 
$E_8$ &  28  & EXT  & $SU(2)_{28} \subset (G_{2})_1$ \\ 
\hline 
\end{tabular}
\label{15} 
\eeq
where $E_6, E_7, E_8$ and $G_2$ are the exceptional Lie groups, while EXT and AUT stand for an extension of the chiral algebra and an automorphism of the theory, respectively. Note that some theories, e.g., $k=16$ have a $D$ as well as a $E$ invariant, as case that we will now discuss in detail.

\subsection{SU(2)$_{16}$}

The $\mathrm{SU}(2)_{16}$ CFT is special in that it has two different off-diagonal partition functions, given by
[recall Eq.~(\ref{ad1})]
\begin{equation}
\begin{split}
Z_{D_{10}}=&
\left|\chi_0+\chi_{16}\right|^2
+\left|\chi_2+\chi_{14}\right|^2
+\left|\chi_4+\chi_{12}\right|^2
\\
&
+\left|\chi_6+\chi_{10}\right|^2
+2\left|\chi_8\right|^2
\end{split}
\end{equation}
and 
\begin{equation}
\begin{split}
Z_{E_7}=&
\left|\chi_0+\chi_{16}\right|^2
+(\chi_2+\chi_{14})\bar{\chi}_{8}
+\chi_{8}(\bar{\chi}_2+\bar{\chi}_{14})
\\&
+\left|\chi_4+\chi_{12}\right|^2
+\left|\chi_6+\chi_{10}\right|^2
+\left|\chi_8\right|^2.
\end{split}
\end{equation}
Both of these theories correspond to a condensation of the boson $a=16$.
There are exactly two distinct solutions to the equation $[M,S]=[M,T]=0$,
given by
\begin{equation}
M^{(1)}=nn^{\mathsf{T}}
,\qquad
M^{(2)}=n\tilde{P}n^{\mathsf{T}},
\end{equation}
where
\begin{equation}
n^{\mathsf{T}}=
\left(
\begin{array}{ccccccccccccccccc}
 1 & 0 & 0 & 0 & 0 & 0 & 0 & 0 & 0 & 0 & 0 & 0 & 0 & 0 & 0 & 0 & 1 \\
 0 & 0 & 1 & 0 & 0 & 0 & 0 & 0 & 0 & 0 & 0 & 0 & 0 & 0 & 1 & 0 & 0 \\
 0 & 0 & 0 & 0 & 1 & 0 & 0 & 0 & 0 & 0 & 0 & 0 & 1 & 0 & 0 & 0 & 0 \\
 0 & 0 & 0 & 0 & 0 & 0 & 1 & 0 & 0 & 0 & 1 & 0 & 0 & 0 & 0 & 0 & 0 \\
 0 & 0 & 0 & 0 & 0 & 0 & 0 & 0 & 1 & 0 & 0 & 0 & 0 & 0 & 0 & 0 & 0 \\
 0 & 0 & 0 & 0 & 0 & 0 & 0 & 0 & 1 & 0 & 0 & 0 & 0 & 0 & 0 & 0 & 0 \\
\end{array}
\right)
\end{equation}
and
\begin{equation}
\tilde{P}
=
 \left(
\begin{array}{cccccc}
 1 & 0 & 0 & 0 & 0 & 0 \\
 0 & 0 & 0 & 0 & 0 & 1 \\
 0 & 0 & 1 & 0 & 0 & 0 \\
 0 & 0 & 0 & 1 & 0 & 0 \\
 0 & 0 & 0 & 0 & 1 & 0 \\
 0 & 1 & 0 & 0 & 0 & 0 \\
\end{array}
\right)
\label{D10}
\end{equation}
is an automorphism of the theory $\mU$.
These two solutions for $M$ are in one-to-one correspondence with the two off-diagonal partition functions above.
Here, $M^{(1)}$ encodes the condensation transition itself, while the existence of the additional matrix $M^{(2)}$ is related to the ``naturality theorem'' discussed in the main text and in Appendix~\ref{sec.chiralextension}.
 
Interestingly, the equation 
$Sn=n\tilde{S}$, that yields the $S$ matrix of the theory after condensation, has two distinct solutions $\tilde{S}$ 
and $\tilde{S}'$, where
\begin{equation}
\tilde{S}
=
\frac{2}{3}
\left(
\begin{array}{cccccc}
 \sin \left(\frac{\pi }{18}\right) & \frac12&  \cos \left(\frac{2 \pi }{9}\right) &  \cos \left(\frac{\pi }{9}\right) & \frac12 & \frac12 \\
 \frac12 & 1 & \frac12 & -\frac12 & -\frac12 & -\frac12 \\
  \cos \left(\frac{2 \pi }{9}\right) & \frac12 & - \cos \left(\frac{\pi }{9}\right) & -\sin \left(\frac{\pi }{18}\right) & \frac12 & \frac12 \\
  \cos \left(\frac{\pi }{9}\right) & -\frac12 & - \sin \left(\frac{\pi }{18}\right) &  \cos \left(\frac{2 \pi }{9}\right) & -\frac12 & -\frac12 \\
 \frac12& -\frac12& \frac12 & -\frac12 & -\frac12 & 1 \\
 \frac12 & -\frac12 & \frac12 & -\frac12 & 1 & -\frac12 \\
\end{array}
\right)
\end{equation}
and $\tilde{S}'$ is obtained by exchanging the last two rows of $\tilde{S}$, a so-called Galois symmetry~\cite{Boer91}. Both  matrices $\tilde{S}$ and $\tilde{S}'$ yield the same fusion rules $\tilde{N}_t$, e.g.,
\begin{equation}
\tilde{N}_{\tilde{3}}=
\left(
\begin{array}{cccccc}
 0 & 0 & 0 & 1 & 0 & 0 \\
 0 & 0 & 1 & 1 & 1 & 1 \\
 0 & 1 & 1 & 2 & 1 & 1 \\
 1 & 1 & 2 & 2 & 1 & 1 \\
 0 & 1 & 1 & 1 & 0 & 1 \\
 0 & 1 & 1 & 1 & 1& 0 \\
\end{array}
\right).
\label{eq: N tilde 3 of SU2 level 16}
\end{equation}
It is worth noting that this condensation of a theory without multiplicities (all $N_{ab}^c$
in SU(2)$_{16}$ are 0 or 1) yields a theory with multiplicities: some of the  $\tilde{N}_{tr}^s$ in Eq.~\eqref{eq: N tilde 3 of SU2 level 16} are larger than 1.

\subsection{SU$(2)_{28}$}

The SU(2)$_{28}$ CFT is special in that it also has two different off-diagonal partition functions, given by
[recall Eq.~(\ref{ad1})]
\begin{equation}
\begin{split}
Z_{D_{16}}=&
\left|\chi_0+\chi_{28}\right|^2
+\left|\chi_2+\chi_{26}\right|^2
+\left|\chi_4+\chi_{24}\right|^2
\\
&
+\left|\chi_6+\chi_{22}\right|^2
+\left|\chi_8+\chi_{20}\right|^2
+\left|\chi_{10}+\chi_{18}\right|^2
\\
&
+\left|\chi_{12}+\chi_{16}\right|^2
+2\left|\chi_{14}\right|^2
\end{split}
\end{equation}
and 
\begin{equation}
\begin{split}
Z_{E_8}=&
\left|\chi_0+\chi_{10}+\chi_{18}+\chi_{28}\right|^2
\\&+
\left|\chi_6+\chi_{12}+\chi_{16}+\chi_{22}\right|^2.
\end{split}
\end{equation}
As is clear from $Z_{E_8}$, the particles 10, 18, and 28 are bosons, besides the vacuum 0. 
The two partition functions correspond to two distinct condensations possible in SU(2)$_{28}$. These are the only condensations possible. Their corresponding $n$ matrices can be read off directly from these partition functions.

The partition function $Z_{E_8}$ stands for a condensation of all bosons with $n^\varphi_a=1$ each. The resulting theory is the Fibonacci TQFT with particles 6, 12, 16, and 22 each restricting to the $\tau$ particle.

The partition function $Z_{D_{16}}$ corresponds to the condensation of the top-level boson 28 only, which results in a 9-particle non-Abelian TQFT with some multiplicities larger than $1$. For example, the restriction of 10 and 18, which we call $\tilde{5}$, obeys the fusion rule [c.f. \eqnref{eq: N tilde 3 of SU2 level 16}]
\begin{equation}
\tilde{5}\times\tilde{5}=\tilde{0}+\tilde{1}+\tilde{2}+\tilde{3}+2\cdot\tilde{4}+2\cdot\tilde{5}+2\cdot\tilde{6}+\tilde{7}+\tilde{8}.
\end{equation}

\section{Proof of Eqs.~\eqref{eq: DU from beta} and~\eqref{eq: DA from beta}}
\label{app.lemma}
We can show Eq.~\eqref{eq: DU from beta} via
\begin{equation}
\begin{split}
\sum_{t\in\mU}\beta_t\beta_t^* 
&=\sum_{t\in\mU}\sum_{a,b\in\mA} d_a d_b n^t_a n^t_b \theta_a \theta_b^*\\
&=\sum_{a,b\in\mA} d_a d_b \sum_{t\in\mU} n^t_a n^t_b\\
&= \sum_{t\in\mU} \left(\sum_{a\in\mA} d_a n^t_a\right)^2\\
&= q^2\sum_{t\in\mU} d_t^2 \\
&= q^2 D_\mU^2,
\end{split}
\end{equation}
where in the second equality, we have used $\theta_a=\theta_b,~\forall~n^t_an^t_b\neq 0$, when $t\in\mU$, because in this case both $a$ and $b$ are in the lift of a deconfined $t$. 

We can show Eq.~\eqref{eq: DA from beta} via
\begin{equation}
\begin{split}
\sum_{t\in\mT}\beta_t\beta_t^*
=&\sum_{a,b\in\mA} d_a d_b \theta_a \theta_b^* \sum_{t\in\mT} n^t_a n^t_b\\
=&\sum_{a,b,c\in\mA} d_a d_b \theta_a\theta_b^* N^b_{ac} n^\varphi_c\\
=&\sum_{b,c\in\mA} d_b n^\varphi_c \sum_{a\in\mA} d_a \theta_a\theta_b^* N^a_{c\bar{b}}\\
=& D_\mA \sum_{b,c\in\mA} d_b n^\varphi_c \theta_c S_{cb}\\
=& D_\mA^2 \sum_{c\in\mA} \left(\sum_{b\in\mA}S_{cb}S_{b1}\right) n^\varphi_c\theta_c\\
=& D_\mA^2 \sum_{c\in\mA} \delta_{c1} n^\varphi_c\theta_c	\\
=& D_\mA^2.
\end{split}
\end{equation}

Moreover, we can show another relation that is useful in Appendix~\ref{app.beta}
\begin{equation}
\label{eq.betabetaN}
\begin{split}
\sum_{r,s\in\mT} \beta_r \beta_s^* \tilde{N}_{rs}^t	
=& \sum_{a,b\in\mA} d_a d_b \frac{\theta_{a}}{\theta_b} \sum_{r,s\in\mT} n_a^r n_b^s \tilde{N}_{rs}^t \\
=& \sum_{a,b,c\in\mA} d_a d_b \frac{\theta_{a}}{\theta_b} N_{ab}^c n_c^t	\\
=& D_\mA^2 \sum_{c,b\in\mA} \theta_c S_{1b} S_{bc} n_c^t	\\
=& D_\mA^2 n_1^t	\\
=& D_\mA^2 \delta_{t, \varphi}.
\end{split}
\end{equation}

\section{$\beta_t =0$ for the simplest case of condensation}
\label{app.beta}

In this section, we prove that $\beta_t=0$ for a confined particle $t\in\mT/\mU$ in a special case: there is only one confined particle $t_0$ which has two lifts with lifting coefficients $1$. For clarity, we start by listing the assumptions used in the following proof. Several of them have been emphasized in the main text. To be complete, we repeat them here.

\begin{enumerate}
\item	$\sum_{r,s\in\mT} \tilde{N}_{rs}^t n_a^r n_b^s = \sum_{c\in\mA} N_{ab}^c n_c^t$.
\item	A deconfined particle $t\in\mU \subset\mT$ has $\theta_a = \theta_b$ if both $n_a^t \neq 0, n_b^t \neq 0$ $\forall a,b$. Otherwise, it is confined. It follows from this definition that every confined particle must have at least two lifts in the $\mA$ theory.
\item	$\tilde{N}_{rs}^t =0$ if $r,s\in\mU$ and $t\in\mT/\mU$. That is, $\mU$ is closed under fusion.
\item	$\mA$ is a MTC.
\item	$\mT/\mU= t_0$. Moreover, $n_{a_1}^{t_0} =n_{a_2}^{t_0} = 1$. For all other $a \ne a_1, a_2 \in\mA$, $n_a^{t_0}= 0$.
\end{enumerate}

\begin{proof}

For clarity, we divide the long proof into several subsections below.

\subsection{Quantum embedding index}

The main task of this section is to prove that in this simplest case, we have $q=2$. That means the condensate has only one nontrivial boson with quantum dimension 1. As the condensate has only one boson, this boson is its own antiparticle. The boson, since it has quantum dimension 1 and is its own anti-particle, is a power 2 simple current (see Sec.~\ref{sec.simplecurrent}), and hence has $\beta_{t_0}=0$ as proved in Sec.~\ref{sec.simplecurrent}. Here we prove the same result in a different way. From Eqs.~\eqref{eq: DU from beta} and \eqref{eq: DA from beta}, we have that
\begin{eqnarray}
\begin{split}
\beta_{t_0} \beta_{t_0}^* &= 
\sum_{t\in\mT}
 \beta_{t} \beta_{t}^*
-\sum_{t\in\mU} 
\beta_{t} \beta_{t}^*	\\
&= D_\mA^2 - q^2 D_\mU^2 	\\
&= D_\mA^2 (1-q) + q^2 d_{t_0}^2,
\end{split}
\label{betat0betat0cc1}
\end{eqnarray}
where we used that $D_\mU^2+ d_{t_0}^2 = D_\mT^2 = D_\mA^2/q$. Put in another way,
\begin{eqnarray}
2 \sum_{a<b\in\mA} n_a^{t_0} n_b^{t_0} d_a d_b [1- \cos(\alpha_a - \alpha_b)] = D_\mA^2(q-1),\label{quantumdimensionofonet0}
\end{eqnarray}
where $\theta_a= \exp(\mathrm{i} \alpha_a)$. This equation imposes quite strong constraints because the righthand side equals $ (1+ x + \sum_{a\in\mA} d_a^2 \delta_{n_a^{t_0} >0})(q-1)$, where the $x\ge 0$ is the cumulative quantum dimension squared of the particles of $\mA$ that split into the deconfined particles of the $\mU$ and that are not identity. 
For our case of only $n_{a_1}^{t_0}=n_{a_2}^{t_0}=1$ the above equation gives
\begin{equation}
2 d_{a_1}d_{a_2} \left[1- \cos(\alpha_{a_1}- \alpha_{a_2})\right] \ge (1+ d_{a_1}^2 + d_{a_2}^2) (q-1).
\end{equation}
For $q\ge3$, this inequality cannot hold. Hence we arrive at the conclusion that for a theory with only one confined particle in $\mT/\mU$ that lifts to only two particles in $A$ with unit lifting coefficients $n$, there can be \emph{only} one condensed boson, and hence $2\le q < 3$. 

However, $2\leq q<3$ implies $q=2$ by the following reason: using the definition of $q$, the inequality can be rewritten
\begin{equation}
 1 \le \sum_{b\in\mA;b\ne 1} n_b^\varphi d_b < 2,
\end{equation} 
where the $1$ excluded in the summand is the vacuum in $\mA$, and $\varphi$ is the vacuum of the $\mT$ theory. Hence there is one and only one boson $B$ with $n_B^\varphi=1$ and $1\le d_B<2$. However, if its quantum dimension is smaller than 2, it cannot split since
\begin{equation}
 d_B = 1 + \sum_{t\neq\varphi\in\mT} n_B^t d_t	
\end{equation} 
would imply  
\begin{equation}
\sum_{t\neq\varphi\in\mT} n_B^t d_t < 1,
\end{equation}
which has only the solution $n_B^t=0$ for all $t\neq\varphi\in\mT$, because $d_t\geq1$.

Hence there is only one condensed boson, and it has quantum dimension $1$, and moreover  $q=2$. Thus $B$ is a simple current which implies $\beta_{t_0}=0$ by the proof in Sec.~\ref{sec.simplecurrent}. We have obtained this result in a different way that also reveals other general properties of theories satisfying the assumptions 1--5. For $q=2$, using \eqnref{quantumdimensionofonet0}, we then have
\begin{equation}
\begin{split}
D_\mA^2 &= 2 d_{a_1}d_{a_2}[1- \cos(\alpha_{a_1}- \alpha_{a_2})]	\\
&= 1+x+ d_{a_1}^2+ d_{a_2}^2.
\end{split}
\label{totalQuantumdimensionofAtheory1}
\end{equation}
Since $x\ge 0$ we necessarily have $\pi/2 < \alpha_{a_1}- \alpha_{a_2} <3 \pi/2$. We now prove that $d_{t_0}= d_{a_1} = d_{a_2}$ as follows. Summing 
\begin{equation}\label{eq:qdimai}
d_{a_i} = d_{t_0} + \sum_{t \ne t_0} n_{a_i}^{t}  d_{t}
\end{equation}
 over $i=1,2$ yields
\begin{equation} 
d_{a_1} + d_{a_2} = 2 d_{t_0} + \sum_{t \ne t_0} (n_{a_1}^{t} +n_{a_2}^{t})d_{t}.
\end{equation}
Further, since $q d_{t_0} = d_{a_1}+ d_{a_2} = 2d_{t_0}$ [from \eqnref{eq.qtdimTA}], it follows that
\begin{equation}
\begin{split} 
& \sum_{t \ne t_0} ( n_{a_1}^{t} +n_{a_2}^{t}  )d_{t}=0,
\end{split}
\end{equation}
which implies $ n_{a_1}^{t} = n_{a_2}^{t} =0,\  \forall t \ne t_0$.
Hence from \eqnref{eq:qdimai}
\begin{eqnarray}
d_{t_0}= d_{a_1} = d_{a_2}.
\end{eqnarray}

To summarize, using assumptions 1--5, we have proved that $q=2$ and the condensate has only one nontrivial boson $B$ with quantum dimension 1. Moreover, the quantum dimensions of $t_0$, $a_1$, and $a_2$ are the same, $d_{t_0}=d_{a_1}=d_{a_2}$. 

\subsection{Fusion rules}

In this section, we find the fusion properties of $a_1$ and $a_2$. Recall from the previous section that these two particles restrict only to the confined particle $t_0$. From \eqnref{eq.commuteRstrFus} we have
\begin{eqnarray}
\sum_{t\in\mT} n_a^t n_b^t = \sum_{c\in\mA} N_{ac}^b n_c^\phi = \delta_{a,b} + N_{aB}^b.
\end{eqnarray} 
Now choose $a=a_1$ and $b \ne a_1, a_2$. The left-hand side is then zero, as $a_1$ only goes into the confined particle $t_0$ while no other particle in $\mA$ besides $a_1, a_2$ restrict to $t_0$ (assumption 5). It follows that
\begin{eqnarray}
0 = N_{a_1 B}^b,~b\ne a_1, a_2.
\end{eqnarray} 
By choosing  $b=a_2$, we have
\begin{equation}
1=n^{t_0}_{a_1}n^{t_0}_{a_2}= N_{a_1 B}^{a_2}
\end{equation}
and similarly for $a_1$ and $a_2$ interchanged. We hence proved that:
\begin{eqnarray}
\begin{split}
a_1\times B &= a_2, 	\\
a_2\times B &= a_1, \\
a_1\times a_1 &= 1 +...\ ,	\\
a_2\times a_2 &= 1 +...\ .
\end{split}
\end{eqnarray} 
By comparing the quantum dimension, we have exhausted all fusion channels in the first two equations. Also, $\bar{a}_1 = a_1$, $\bar{a}_2 = a_2$, because if $\bar{a_1} = a_2$ then $\theta_{a_1} = \theta_{a_2}$, which is not allowed since $t_0$ is by assumption confined.

Using the quantum dimension for the $\mT$ theory $\sum_{t\in\mT} N_{rs}^t d_t = d_r d_s$, we have the following implications:
\begin{equation}
\begin{split}
r,s \in\mU &\Rightarrow \sum_{t\in\mU} \tilde{N}_{rs}^t d_t = d_r d_s, \\ 
r \in\mT/\mU = t_0, s\in\mU &\Rightarrow t\in\mT/\mU=t_0 \\
&\Rightarrow \tilde{N}_{t_0s}^{t_0} d_{t_0} = d_{t_0} d_s,	\\
&\Rightarrow \tilde{N}_{t_0 s}^{t_0} = d_s, \\ 
r,s\in\mT/\mU = t_0 &\Rightarrow \tilde{N}_{t_0 t_0}^{t_0} d_{t_0} + \sum_{t\in\mU} \tilde{N}_{t_0 t_0}^t d_t = d_{t_0}^2, \\
&\Rightarrow \tilde{N}_{t_0 t_0}^{t_0} d_{t_0} + \sum_{t\in\mU} d^t d_t = d_{t_0}^2, \\
&\Rightarrow \tilde{N}_{t_0 t_0}^{t_0} d_{t_0} +D_\mU^2 = d_{t_0}^2.
\end{split}
\label{quantumdimensionsofhteTtheory1}
\end{equation}

The last equation together with $D_\mT^2= D_\mU^2 + d_{t_0}^2$ gives
\begin{eqnarray}\label{eq:b14}
d_{t_0} (2 d_{t_0} - \tilde{N}_{t_0 t_0}^{t_0}) = D_\mT^2.
\end{eqnarray}

From Eq.~\eqref{quantumdimensionsofhteTtheory1}, since $\tilde{N}^{t_0}_{t_0s}=d_s$, every deconfined particle $s$ has to appear in the fusion of the confined particle with itself and the quantum dimension of every deconfined particle is an integer.

We now refine the statement as a summary of this section: for a theory with only one confined particle $t_0\in \mT/\mU$ which lifts to only two particles $a_1$, $a_2$ in $\mA$ with unit lifting coefficients $n_{a_1}^{t_0} = n_{a_2}^{t_0} $, there can be \emph{only} one condensed boson of quantum dimension $1$, $d_{t_0} = d_{a_1} = d_{a_2}$ and $\tilde{N}_{t_0 t_0}^r = d_r \in Z_+$,  $\forall r \in\mU$. Also, $a_1 \times B= a_2$, and furthermore $a_1 = \bar{a}_1$, $a_2= \bar{a}_2$.

\subsection{$\beta_{t_0}=0$}

We now use \eqnref{eq: intermediate step in nS proof 0} for $t= t_0$ and expand the sum using the assumption that two deconfined particles cannot fuse to a confined particle
\begin{equation}
\begin{split}
\theta_b D_\mA &\sum_{c\in\mA} S_{bc} \theta_c n_c^{t_0}	\\
=&\left(\sum_{r,s\in\mT/\mU} + \sum_{r\in\mT/\mU} \sum_{s\in\mU} +  \sum_{s\in\mT/\mU} \sum_{r\in\mU}\right) \tilde{N}_{rs}^{t_0} n_b^s \beta_r	\\ 
=& \tilde{N}_{t_0 t_0}^{t_0} n_b^{t_0} \beta_{t_0} + \sum_{s\in\mU} \tilde{N}_{t_0 s}^{t_0} n_b^s \beta_{t_0} + \sum_{r\in\mU} \tilde{N}_{r t_0 }^{t_0} n_b^{t_0} \beta_{r}	\\
=& \tilde{N}_{t_0 t_0}^{t_0} n_b^{t_0} \beta_{t_0} + \sum_{s\in\mU} d_s n_b^s \beta_{t_0} + \sum_{r\in\mU} d_r n_b^{t_0} \beta_{r}   \\ 
=& \tilde{N}_{t_0 t_0}^{t_0} n_b^{t_0} \beta_{t_0} + \sum_{s\in\mT} d_s n_b^s \beta_{t_0} + \sum_{r\in\mT} d_r n_b^{t_0} \beta_{r} \\
&- d_{t_0} n_b^{t_0} \beta_{t_0} - d_{t_0} \beta_{t_0} n_b^{t_0}	\\ 
=& \beta_{t_0} n_b^{t_0} (\tilde{N}_{t_0 t_0}^{t_0} - 2 d_{t_0}) + \sum_{s\in\mT} d_s n_b^s \beta_{t_0} + \sum_{r\in\mT} d_r n_b^{t_0} \beta_{r}	\\
=& - \beta_{t_0} n_b^{t_0}\frac{D_\mT^2}{d_{t_0}} + d_b \beta_{t_0} + \sum_{r\in\mT} d_r n_b^{t_0} \beta_{r},
\end{split}
\end{equation}
where we have used Eq.~\eqref{quantumdimensionsofhteTtheory1} and in the last line we used Eq.~\eqref{eq:b14}. We can compute the remaining sum easily:
\begin{equation}
\begin{split}
\sum_{r\in\mT} d_r \beta_r &= \sum_{r\in\mT} \sum_{a\in\mA} d_r n_a^r d_a \theta_a	\\
&= \sum_{a\in\mA} d_a^2 \theta_a	\\
&= D_\mA \Theta_{\mA},
\end{split}
\end{equation} 
where $\Theta_{\mA} = \exp(\mathrm{i} 2 \pi c/8)$ with $c$ the central charge of the $\mA$ theory. We have hence proved the relation
\begin{eqnarray}
\begin{split}
&\theta_b D_\mA \sum_{c \in\mA} S_{bc} \theta_c n_c^{t_0}=
\beta_{t_0} \left(d_b - \frac{D_\mT^2}{d_{t_0}}n_b^{t_0}\right) + D_\mA \Theta_\mA n_b^{t_0}.
\end{split}
\end{eqnarray}
For the case when the particle $t_0$ has two lifts with coefficients $1$, we have
\begin{eqnarray}
\begin{split}
&\theta_b D_\mA  (S_{ba_1} \theta_{a_1} + S_{ba_2} \theta_{a_2}) \\
&\quad= \beta_{t_0} \left(d_b - \frac{D_\mT^2}{d_{t_0}}n_b^{t_0}\right) + D_\mA \Theta_\mA n_b^{t_0}.
\end{split}
\label{zerothetaequation}
\end{eqnarray}
Note that the righthand side of the above equation is the same whether we choose $b= a_1$ or $b=a_2$ -- use  $d_{t_0} = d_{a_1} = d_{a_2}$. Therefore, 
\begin{equation}
\begin{split}
S_{a_1 a_1} \theta_{a_1}^2 + S_{a_1 a_2} \theta_{a_1} \theta_{a_2}   &= S_{a_2 a_2} \theta_{a_2}^2 + S_{a_1 a_2} \theta_{a_1} \theta_{a_2}	\\
\Rightarrow~~ S_{a_1 a_1} \theta_{a_1}^2 &= S_{a_2 a_2} \theta_{a_2}^2.
\end{split}
\label{firstthetaequation}
\end{equation}

We now prove another relation. Starting from \eqnref{eq.commuteRstrFus} and using the Verlinde formula for MTCs gives
\begin{equation}
\begin{split}
 \sum_{s,r\in\mT} \tilde{N}_{rs}^t n_a^r n_b^s = &\,\sum_{c\in\mA} N_{ab}^c n_c^t	\\
=&\, \sum_{c,m\in\mA}S_{bm}\frac{S_{am}}{S_{1m}} S_{\bar{c} m} n_c^t	\\
=&\, \sum_{m\in\mA} S_{bm}\frac{S_{am}}{S_{1m}} \sum_{c\in\mA} S_{\bar{c} m} n_c^t .
\end{split} 
\end{equation} 
Multiply both sides by $S_{\bar{b}e}$ and sum over $b$, while using
\begin{eqnarray}
\sum_{b\in\mA} S_{\bar{b}e} S_{bm} = \sum_{b\in\mA} S_{b\bar{e}} S_{bm} = \sum_{b\in\mA} S_{\bar{e}b} S_{bm} = \delta_{e m},
\end{eqnarray} 
where we have used the property of $S$ matrix $S_{ab}=S_{\bar{a}\bar{b}}$. This gives 
\begin{eqnarray}
\sum_{s,r\in\mT} \tilde{N}_{rs}^t n_a^r \sum_{b\in\mA} S_{\bar{b}e} n_b^s = \frac{S_{ae}}{S_{1e}} \sum_{c \in\mA} S_{\bar{c} e} n_c^t.
\end{eqnarray} 
Defining $P_a^t \equiv \sum_{c\in\mA} S_{ca} n_c^t$, we rewrite
\begin{eqnarray}
\sum_{s,r\in\mT} \tilde{N}_{\bar{r}s}^{t} n_{a}^r P_e^s= \frac{S_{ae}}{S_{1e}} P_e^t.
\end{eqnarray}
We now pick $a=B$, the condensed boson. Since $d_B=1$ as we have proved, we only have nonzero $n_B^\varphi=1$, and we have
\begin{eqnarray}
\sum_{s\in\mT} \tilde{N}_{\varphi s}^{t} P_e^s =(P_e^t) = \frac{S_{Be}}{S_{1e}} P_e^t.
\end{eqnarray} 
There are hence two solutions: 

\textbf{A}. $P_e^t =0$

\textbf{B}. $S_{Be}/S_{1e}=1$ if $P_e^t\neq 0$, $\forall e\in\mA,~\forall t\in\mT$. 

For $e = a_1, a_2$ case \textbf{B} is not possible. We show this as follows: Assume $e$ equals either $a_1$ or $a_2$.  Then \textbf{B} implies:
\begin{equation}
\begin{split}
\sum_{c\in\mA} N_{B a_1}^c \theta_c d_c =& \sum_{c\in\mA} N_{1 a_1}^c \theta_c d_c	\\
=& \,\theta_{a_1} d_{a_1}=\theta_{a_2} d_{a_2},
\end{split}
\end{equation} 
where we used Eq.~\eqref{eq: def S} which is an expression for the entries of $S$ matrix, and $N_{B a_1}^b = \delta_{b a_2}$. Since $d_{a_1} = d_{a_2}$ this would imply $\theta_{a_1}=\theta_{a_2}$, which again is not possible as $t_0$ is confined. 

Hence for $e$ equals either $a_1$ or $a_2$ we have:
\begin{eqnarray}
P_{a_1}^t=P_{a_2}^t =0,~~\forall t\in\mT.
\end{eqnarray} 
Choosing $t=t_0$ we have the two equalities
\begin{equation}
S_{a_1 a_1}+ S_{a_2 a_1} = S_{a_2 a_2} + S_{a_2 a_1} =0
\end{equation} 
which imply
\begin{equation}
S_{a_1 a_1}= S_{a_2 a_2}.
\end{equation}
This equation, along with \eqnref{firstthetaequation} now has the two solutions
\begin{eqnarray}
\text{\textbf{A1}:~~} S_{a_1 a_1}= S_{a_2 a_2} = S_{a_1 a_2}=0 \label{firstproblemequation}
\end{eqnarray}
and
\begin{eqnarray}
\text{\textbf{A2}:~~} \theta_{a_1}^2 = \theta_{a_2}^2.
\end{eqnarray}

If \textbf{A2} is the solution, we are done because it means $\theta_{a_1} = - \theta_{a_2}$ (the equality $\theta_{a_1} = \theta_{a_2}$ is not possible by assumption of $t_0$ confined) and hence $\beta_{t_0}=0$, which is what we wanted to prove.

For \textbf{A1}, we can prove that it does not yield a consistent solution.
Observe that \eqnref{zerothetaequation} gives
\begin{eqnarray}
\beta_{t_0} \left(d_{a_1} - \frac{D_\mA^2}{2 d_{a_1}}\right) = - D_\mA \Theta_\mA\, ,
\end{eqnarray} 
where we have used $D_\mT^2= D_\mA^2/q= D_\mA^2/2$. By definition $\beta_{t_0} = d_{a_1} (\theta_{a_1} + \theta_{a_2})$. This gives the equation:
\begin{eqnarray}
e^{\mathrm{i}(\alpha_1 - \frac{2 \pi c}{8})}+ e^{\mathrm{i}(\alpha_2 - \frac{2 \pi c}{8})} = - \frac{2 D_\mA}{2 d_{a_1}^2 - D_\mA^2}.
\end{eqnarray} 
Since the righthand side is real, we must have $\alpha_1 - \frac{2\pi c}{8}=-(\alpha_2-\frac{2\pi c}{8}) = \gamma_1$ and hence
\begin{eqnarray}
\cos \gamma_1 = - \frac{ D_\mA}{2 d_{a_1}^2 - D_\mA^2}.\label{gamma1andDaequation}
\end{eqnarray} 

Using \eqnref{totalQuantumdimensionofAtheory1} we obtain $D_\mA^2 =2 d_{a_1}^2 [1- \cos(2 \gamma_1)] = 4 d_{a_1}^2 \sin(\gamma_1)^2$  (with $ \pi/2< 2 \gamma_1< 3 \pi/2$, since $D_\mA^2\ge 1+ 2 d_{a_1}^2$) and substituting this in \eqnref{gamma1andDaequation} becomes
\begin{eqnarray}
\cos(\gamma_1) = - \frac{|\sin(\gamma_1)|}{d_{a_1} \cos(2 \gamma_1)}.
\end{eqnarray} 
We also have:
\begin{equation}
\begin{split}
D_\mA^2 =& q D_\mT^2 \\
=& 2(d_{t_0}^2+ D_{\mU}^2)	\\
=& 2\left(2 d_{t_0}^2 - \tilde{N}_{t_0 t_0}^{t_0} d_{t_0}\right)	\\
=& 4 d_{a_1}^2 - 2 \tilde{N}_{t_0 t_0}^{t_0} d_{a_1} ,
\end{split}
\end{equation}
where we have used that $d_{t_0} = d_{a_1}$ as well as \eqnref{quantumdimensionsofhteTtheory1}. Plugging in the expression for $D_\mA$ we then have
\begin{eqnarray}
\tilde{N}_{t_0 t_0}^{t_0} = d_{a_1} [ 1+ \cos(2 \gamma_1)] = 2 d_{a_1} \cos(\gamma_1)^2.
\end{eqnarray} 
We need one last equation to show that case \textbf{A1} cannot be true. This comes from \eqnref{eq.betabetaN} which we use for the case $t= t_0$
\begin{eqnarray}
\tilde{N}_{t_0 t_0}^{t_0} \beta_{t_0} \beta_{t_0}^* + \sum_{s\in\mU} \tilde{N}_{t_0 s}^{t_0} \beta_s \beta_{t_0}^* + \sum_{s\in\mU} \tilde{N}_{t_0 s}^{t_0} \beta_s^* \beta_{t_0} =0.
\end{eqnarray} 
From the already proved $\tilde{N}_{t_0 s}^{t_0}=d_s$ and $\sum_{s\in\mT} d_s \beta_s = D_\mA \Theta_\mA$ we find:
\begin{equation}
\begin{split}
0=&\beta_{t_0}\beta_{t_0}^* (\tilde{N}_{t_0 t_0}^{t_0} - 2 d_{t_0}) + \beta_{t_0}^* D_\mA \Theta_\mA +  \beta_{t_0} D_\mA \Theta_\mA^*	\\ 
=& - \beta_{t_0}\beta_{t_0}^* \frac{D_\mT^2}{d_{t_0}} + \beta_{t_0}^* D_\mA \Theta_\mA +  \beta_{t_0} D_\mA \Theta_\mA^*	\\
=& - (4 d_{t_0}^2 - D_\mA^2) \frac{D_\mA^2}{q d_{t_0}}  +\beta_{t_0}^* D_\mA \Theta_\mA +  \beta_{t_0} D_\mA \Theta_\mA^*,
\end{split}
\end{equation} 
where we have used that $ \beta_{t_0}\beta_{t_0}^* =  4 d_{t_0}^2 - D_\mA^2 $,
which follows from Eq.~\eqref{betat0betat0cc1} with $q=2$.
Hence the above becomes
\begin{equation}
\begin{split}
0=& - \tilde{N}_{t_0 t_0}^{t_0}D_\mA  +\beta_{t_0}^* \Theta_\mA +  \beta_{t_0} \Theta_\mA^*	\\
=& - \tilde{N}_{t_0 t_0}^{t_0}D_\mA+ 2 d_{a_1}[ \cos(\gamma_1) + \cos(\gamma_2)]	\\
=& - \tilde{N}_{t_0 t_0}^{t_0}D_\mA+ 4 d_{a_1} \cos(\gamma_1).
\end{split}
\end{equation} 
Since $\pi/2< \gamma_1< 3 \pi/2$, the righthand side is negative so the equation cannot possibly hold. Hence only case \textbf{A2} is possible which means $\theta_{a_1} = - \theta_{a_2}$ and hence $\beta_{t_0}=0$.  
\end{proof}

\section{Condensing four layers of Ising TQFT}
\label{sec: 4 layers Ising}

Here we give some detail on one particular condensation in a theory composed of a tensor product of 4 layers of the Ising TQFT. We show that, generically, in our algorithm in the main text \secref{sec.examples}, we do need to check that the resulting $\tilde{S}$ matrix gives integer fusion matrices using Verlinde's formula. This is not a complete discussion of all possible condensations in the 4 layer Ising theory. 

We focus on the condensate containing $1111$, $11\psi\psi$, $1\psi1\psi$, and $1\psi\psi1$, but not $\psi\psi\psi\psi$, $\psi\psi11$, $\psi1\psi1$, and $\psi11\psi$. (Including these other bosons in the condensate would yield the $\nu=4$ theory from Kitaev's 16-fold way.) 
The corresponding $M$ matrix has only one nonzero entry on the column (and row) that corresponds to the $\sigma\sigma\sigma\sigma$ particle, namely $M_{\sigma\sigma\sigma\sigma,\sigma\sigma\sigma\sigma}=4$. Since the quantum dimension $d_{\sigma\sigma\sigma\sigma}=4$, this allows for two distinct solutions $n$: one in which $\sigma\sigma\sigma\sigma$ particle restricts to twice some particle $a$ and one in which it splits into 4 Abelian particles $a_1$, $a_2$, $a_3$, $a_4$.
In both cases, we can find solutions to \eqnref{eq: nS nT equations} that are unitary and satisfy $\tilde{S}^2=\Theta(\tilde{S}\tilde{T})^3=\tilde{C}$.

However, for the theory in which the $\sigma\sigma\sigma\sigma$ particle splits into 4 particles, the fusion coefficients $\tilde{N}_t$ obtained from \eqnref{eq: new fusion coefficients} are not all nonnegative integers (some of them are $\pm1/2$). This establishes by example that we have to impose that  $\tilde{N}_t$ is nonnegative integer valued, in addition to the other conditions in \secref{sec.examples}. It also shows that in the example at hand, despite the ambiguity in the possible solutions for $n$, the particle content (up to possible automorphisms) of the final theory is fixed by the choice of condensate. Whether this statement is true in general is presently not known to us.

The allowed solution, in which $\sigma\sigma\sigma\sigma$ particle restricts to twice the same particle, is the one we naively expect, upon inspection of the anyons in the condensates by the following argument: all condensed anyons have a vacuum particle 1 in the first layer. Hence the Ising theory in the first layer will be preserved under condensation and the result is a direct product of the $\nu=1$ Ising theory and the $\nu=3$ Ising theory from Kitaev's 16-fold way.
The particle that $\sigma\sigma\sigma\sigma$ twice restricts to is the direct product of a $\nu=1$ Ising $\sigma$ and $\nu=3$ Ising $\sigma$, which we have already proved in Sec.~\ref{subsubsec: Ising} has $n_{(\sigma\sigma\sigma)}^{\sigma'}=2$.

\section{Condensations in the quantum double of $D_2$}
\label{app: condensation of D_2}

To demonstrate the power of our approach to condensation, we will list here all possible condensations in the TQFT corresponding to the quantum double of $D_2$. All the information about this theory including its fusion rules can be found in Ref.~\onlinecite{Propitius}. To tackle this task with a less systematic approach would be a challenge, not only because it contains 22 particles, but also because 10 of them are bosons, leading to a wealth of possible condensates.

\begin{widetext}
In the basis $(1,\bar{1},J_1,J_2,J_3,\bar{J}_1,\bar{J}_2,\bar{J}_3,\chi,\bar{\chi},
\sigma^+_1,\sigma^+_2,\sigma^+_3,\sigma^-_1,\sigma^-_2,\sigma^-_3,
\tau^+_1,\tau^+_2,\tau^+_3,\tau^-_1,\tau^-_2,\tau^-_3)$,
the modular $S$ and $T$ matrices of the theory are given by
\footnotesize
\begin{equation}
S=\frac{1}{8}
\left(
\begin{array}{cccccccccccccccccccccc}
 1 & 1 & 1 & 1 & 1 & 1 & 1 & 1 & 2 & 2 & 2 & 2 & 2 & 2 & 2 & 2 & 2 & 2 & 2 & 2 & 2 & 2 \\
 1 & 1 & 1 & 1 & 1 & 1 & 1 & 1 & -2 & -2 & 2 & 2 & 2 & 2 & 2 & 2 & -2 & -2 & -2 & -2 & -2 & -2 \\
 1 & 1 & 1 & 1 & 1 & 1 & 1 & 1 & 2 & 2 & 2 & -2 & -2 & 2 & -2 & -2 & 2 & -2 & -2 & 2 & -2 & -2 \\
 1 & 1 & 1 & 1 & 1 & 1 & 1 & 1 & 2 & 2 & -2 & 2 & -2 & -2 & 2 & -2 & -2 & 2 & -2 & -2 & 2 & -2 \\
 1 & 1 & 1 & 1 & 1 & 1 & 1 & 1 & 2 & 2 & -2 & -2 & 2 & -2 & -2 & 2 & -2 & -2 & 2 & -2 & -2 & 2 \\
 1 & 1 & 1 & 1 & 1 & 1 & 1 & 1 & -2 & -2 & 2 & -2 & -2 & 2 & -2 & -2 & -2 & 2 & 2 & -2 & 2 & 2 \\
 1 & 1 & 1 & 1 & 1 & 1 & 1 & 1 & -2 & -2 & -2 & 2 & -2 & -2 & 2 & -2 & 2 & -2 & 2 & 2 & -2 & 2 \\
 1 & 1 & 1 & 1 & 1 & 1 & 1 & 1 & -2 & -2 & -2 & -2 & 2 & -2 & -2 & 2 & 2 & 2 & -2 & 2 & 2 & -2 \\
 2 & -2 & 2 & 2 & 2 & -2 & -2 & -2 & 4 & -4 & 0 & 0 & 0 & 0 & 0 & 0 & 0 & 0 & 0 & 0 & 0 & 0 \\
 2 & -2 & 2 & 2 & 2 & -2 & -2 & -2 & -4 & 4 & 0 & 0 & 0 & 0 & 0 & 0 & 0 & 0 & 0 & 0 & 0 & 0 \\
 2 & 2 & 2 & -2 & -2 & 2 & -2 & -2 & 0 & 0 & 4 & 0 & 0 & -4 & 0 & 0 & 0 & 0 & 0 & 0 & 0 & 0 \\
 2 & 2 & -2 & 2 & -2 & -2 & 2 & -2 & 0 & 0 & 0 & 4 & 0 & 0 & -4 & 0 & 0 & 0 & 0 & 0 & 0 & 0 \\
 2 & 2 & -2 & -2 & 2 & -2 & -2 & 2 & 0 & 0 & 0 & 0 & 4 & 0 & 0 & -4 & 0 & 0 & 0 & 0 & 0 & 0 \\
 2 & 2 & 2 & -2 & -2 & 2 & -2 & -2 & 0 & 0 & -4 & 0 & 0 & 4 & 0 & 0 & 0 & 0 & 0 & 0 & 0 & 0 \\
 2 & 2 & -2 & 2 & -2 & -2 & 2 & -2 & 0 & 0 & 0 & -4 & 0 & 0 & 4 & 0 & 0 & 0 & 0 & 0 & 0 & 0 \\
 2 & 2 & -2 & -2 & 2 & -2 & -2 & 2 & 0 & 0 & 0 & 0 & -4 & 0 & 0 & 4 & 0 & 0 & 0 & 0 & 0 & 0 \\
 2 & -2 & 2 & -2 & -2 & -2 & 2 & 2 & 0 & 0 & 0 & 0 & 0 & 0 & 0 & 0 & -4 & 0 & 0 & 4 & 0 & 0 \\
 2 & -2 & -2 & 2 & -2 & 2 & -2 & 2 & 0 & 0 & 0 & 0 & 0 & 0 & 0 & 0 & 0 & -4 & 0 & 0 & 4 & 0 \\
 2 & -2 & -2 & -2 & 2 & 2 & 2 & -2 & 0 & 0 & 0 & 0 & 0 & 0 & 0 & 0 & 0 & 0 & -4 & 0 & 0 & 4 \\
 2 & -2 & 2 & -2 & -2 & -2 & 2 & 2 & 0 & 0 & 0 & 0 & 0 & 0 & 0 & 0 & 4 & 0 & 0 & -4 & 0 & 0 \\
 2 & -2 & -2 & 2 & -2 & 2 & -2 & 2 & 0 & 0 & 0 & 0 & 0 & 0 & 0 & 0 & 0 & 4 & 0 & 0 & -4 & 0 \\
 2 & -2 & -2 & -2 & 2 & 2 & 2 & -2 & 0 & 0 & 0 & 0 & 0 & 0 & 0 & 0 & 0 & 0 & 4 & 0 & 0 & -4 \\
\end{array}
\right),
\end{equation}
\normalsize

\begin{equation}
T=\mathrm{diag}
(1,1,1,1,1,1,1,1,1,-1,1,1,1,-1,-1,-1,\mathrm{i},\mathrm{i},\mathrm{i},-\mathrm{i},-\mathrm{i},-\mathrm{i}).
\end{equation}

This theory contains three automorphisms: One can simultaneously exchange the subscripts $1\leftrightarrow2$, $1\leftrightarrow3$, or $2\leftrightarrow3$ on all anyons that carry such an index.

One obtains 31 solutions for $M$ to the equations \eqnref{eq: M, T commutator}, when only symmetric nonnegative integer $M$ with $M_{1,1}=1$ are allowed and the triangle constraint $M_{a,a}+M_{b,b}\geq 2 M_{a,b}$ is imposed in order to avoid $M$ that involve automorphisms in the $\mU$ theory.
One solution is the identity matrix. Four solutions do not admit a decomposition $M=nn^\mathsf{T}$ with a nonnegative integer matrix $n$.  
The remaining 26 solutions are distinct condensations.
Below we give a complete list of all these possible condensates $n^\varphi_a$ that lead to a consistent TQFT. They are grouped by the type of resulting theory.

\emph{Condensation to trivial theory} ---
The following 6 choices of $n^\varphi_a$ lead to a trivial TQFT (only the vacuum is left). Notice that $n^\varphi_a>1$ in some cases.
\begin{equation}
\begin{array}{lccccccccccccc}
\ \ a=& 1 & \bar{1} & J_1 & J_2 & J_3 & \bar{J}_1 & \bar{J}_2 & \bar{J}_3 & \chi  & \sigma ^+_1 & \sigma ^+_2 & \sigma ^+_3 \\
 \hline
 n^\varphi_a=& 1 & 0 & 1 & 1 & 1 & 0 & 0 & 0 & 2 & 0 & 0 & 0 \\
 n^\varphi_a=&  1 & 1 & 0 & 0 & 0 & 0 & 0 & 0 & 0 & 1 & 1 & 1  \\
 n^\varphi_a=&  1 & 1 & 0 & 0 & 1 & 0 & 0 & 1 & 0 & 0 & 0 & 2 \\
 n^\varphi_a=&  1 & 1 & 0 & 1 & 0 & 0 & 1 & 0 & 0 & 0 & 2 & 0  \\
 n^\varphi_a=&  1 & 1 & 1 & 0 & 0 & 1 & 0 & 0 & 0 & 2 & 0 & 0 \\
 n^\varphi_a=&  1 & 1 & 1 & 1 & 1 & 1 & 1 & 1 & 0 & 0 & 0 & 0 \\
\end{array}
\end{equation}

\emph{Condensation to toric code} ---
The following 7 choices of $n^\varphi_a$ lead to the toric code TQFT defined in \eqnref{eq: Toric code}. \begin{equation}
\begin{array}{lccccccccccccc}
\ \ a=& 1 & \bar{1} & J_1 & J_2 & J_3 & \bar{J}_1 & \bar{J}_2 & \bar{J}_3 & \chi  & \sigma ^+_1 & \sigma ^+_2 & \sigma ^+_3 \\
 \hline
 n^\varphi_a=& 1 & 0 & 1 & 1 & 1 & 0 & 0 & 0 & 0 & 0 & 0 & 0  \\
 n^\varphi_a=& 1 & 1 & 0 & 0 & 0 & 0 & 0 & 0 & 0 & 0 & 0 & 1  \\
 n^\varphi_a=& 1 & 1 & 0 & 0 & 0 & 0 & 0 & 0 & 0 & 0 & 1 & 0  \\
 n^\varphi_a=& 1 & 1 & 0 & 0 & 0 & 0 & 0 & 0 & 0 & 1 & 0 & 0  \\
 n^\varphi_a=& 1 & 1 & 0 & 0 & 1 & 0 & 0 & 1 & 0 & 0 & 0 & 0  \\
 n^\varphi_a=& 1 & 1 & 0 & 1 & 0 & 0 & 1 & 0 & 0 & 0 & 0 & 0  \\
 n^\varphi_a=& 1 & 1 & 1 & 0 & 0 & 1 & 0 & 0 & 0 & 0 & 0 & 0
\end{array}
\end{equation}

\emph{Condensation to double semion} ---
The following 6 choices of $n^\varphi_a$ lead to the double semion TQFT with four abelian particles 
$1$, $b$, $s$, $\tilde{s}$
and the modular matrices
\begin{equation}
S_{\mathrm{DS}}
=\frac12
\begin{pmatrix}
1&1&1&1\\
1&1&-1&-1\\
1&-1&-1&1\\
1&-1&1&-1
\end{pmatrix},\qquad
T_{\mathrm{DS}}=\mathrm{diag}(1,1,\mathrm{i},-\mathrm{i}).
\end{equation}
There are examples where $M_{aa}=4$, but the respective particle $a$ has quantum dimension $2$.
If $a$ splits into 4 particles, Eq.~\eqref{eq.qtdimTA} requires it to have at least quantum dimension $4$. Hence,  the solution to $M=nn^{\mathsf{T}}$ needs to have $n_a^t=2$, ruling out the other option of splitting into 4 particles that was discussed in Appendix~\ref{sec: 4 layers Ising} for the 4 layers of Ising TQFT. 
\begin{equation}
\begin{array}{lccccccccccccc}
\ \ a=& 1 & \bar{1} & J_1 & J_2 & J_3 & \bar{J}_1 & \bar{J}_2 & \bar{J}_3 & \chi  & \sigma ^+_1 & \sigma ^+_2 & \sigma ^+_3 \\
 \hline
 n^\varphi_a=& 1 & 0 & 0 & 0 & 0 & 0 & 0 & 1 & 0 & 0 & 0 & 1 \\
 n^\varphi_a=& 1 & 0 & 0 & 0 & 0 & 0 & 1 & 0 & 0 & 0 & 1 & 0 \\
 n^\varphi_a=& 1 & 0 & 0 & 0 & 0 & 1 & 0 & 0 & 0 & 1 & 0 & 0 \\
 n^\varphi_a=& 1 & 0 & 0 & 0 & 1 & 1 & 1 & 0 & 0 & 0 & 0 & 0 \\
 n^\varphi_a=& 1 & 0 & 0 & 1 & 0 & 1 & 0 & 1 & 0 & 0 & 0 & 0 \\
 n^\varphi_a=& 1 & 0 & 1 & 0 & 0 & 0 & 1 & 1 & 0 & 0 & 0 & 0
\end{array}
\end{equation}

\emph{Condensation to Abelian theories with 16 anyons} ---
The following 7 choices of $n^\varphi_a$ lead to Abelian theories with 16 anyons. The $n^\varphi_a$ are grouped and listed below.
\begin{equation}
\begin{array}{lccccccccccccc}
\ \ a=& 1 & \bar{1} & J_1 & J_2 & J_3 & \bar{J}_1 & \bar{J}_2 & \bar{J}_3 & \chi  & \sigma ^+_1 & \sigma ^+_2 & \sigma ^+_3 \\
 \hline
 n^\varphi_a=& 1 & 0 & 0 & 0 & 0 & 0 & 0 & 1 & 0 & 0 & 0 & 0 \\
 n^\varphi_a=& 1 & 0 & 0 & 0 & 0 & 0 & 1 & 0 & 0 & 0 & 0 & 0 \\
 n^\varphi_a=& 1 & 0 & 0 & 0 & 0 & 1 & 0 & 0 & 0 & 0 & 0 & 0 \\
 \hline
 n^\varphi_a=& 1 & 0 & 0 & 0 & 1 & 0 & 0 & 0 & 0 & 0 & 0 & 0 \\
 n^\varphi_a=& 1 & 0 & 0 & 1 & 0 & 0 & 0 & 0 & 0 & 0 & 0 & 0 \\
 n^\varphi_a=& 1 & 0 & 1 & 0 & 0 & 0 & 0 & 0 & 0 & 0 & 0 & 0 \\
 \hline
 n^\varphi_a=& 1 & 1 & 0 & 0 & 0 & 0 & 0 & 0 & 0 & 0 & 0 & 0 \\
\end{array}
\end{equation}

The first three lines share the same $\tilde{S}$, $\tilde{T}$ matrices since and differ by an automorphism in $\mA$. So do the next three lines. The third group has only one condensing boson $\bar{1}$. The modular matrices for each of the groups are listed as follows:
\begin{subequations}
\begin{equation}
\tilde{S}_{16}^{(1)}= S_{\mathrm{DS}}\otimes S_{\mathrm{DS}}
,\qquad
\tilde{T}_{16}^{(1)}=T_{\mathrm{DS}}\otimes T_{\mathrm{DS}},
\end{equation}

\begin{eqnarray}
\begin{split}
\tilde{S}_{16}^{(2)}&=\frac{1}{4}
\left(
\begin{array}{cccccccccccccccc}
 1 & 1 & 1 & 1 & 1 & 1 & 1 & 1 & 1 & 1 & 1 & 1 & 1 & 1 & 1 & 1 \\
 1 & 1 & 1 & 1 & 1 & 1 & 1 & 1 & -1 & -1 & -1 & -1 & -1 & -1 & -1 & -1 \\
 1 & 1 & 1 & 1 & -1 & -1 & -1 & -1 & 1 & 1 & 1 & 1 & -1 & -1 & -1 & -1 \\
 1 & 1 & 1 & 1 & -1 & -1 & -1 & -1 & -1 & -1 & -1 & -1 & 1 & 1 & 1 & 1 \\
 1 & 1 & -1 & -1 & 1 & 1 & -1 & -1 & -\mathrm{i} & \mathrm{i} & \mathrm{i} & -\mathrm{i} & \mathrm{i} & -\mathrm{i} & -\mathrm{i} & \mathrm{i} \\
 1 & 1 & -1 & -1 & 1 & 1 & -1 & -1 & \mathrm{i} & -\mathrm{i} & -\mathrm{i} & \mathrm{i} & -\mathrm{i} & \mathrm{i} & \mathrm{i} & -\mathrm{i} \\
 1 & 1 & -1 & -1 & -1 & -1 & 1 & 1 & -\mathrm{i} & \mathrm{i} & \mathrm{i} & -\mathrm{i} & -\mathrm{i} & \mathrm{i} & \mathrm{i} & -\mathrm{i} \\
 1 & 1 & -1 & -1 & -1 & -1 & 1 & 1 & \mathrm{i} & -\mathrm{i} & -\mathrm{i} & \mathrm{i} & \mathrm{i} & -\mathrm{i} & -\mathrm{i} & \mathrm{i} \\
 1 & -1 & 1 & -1 & -\mathrm{i} & \mathrm{i} & -\mathrm{i} & \mathrm{i} & 1 & 1 & -1 & -1 & \mathrm{i} & -\mathrm{i} & \mathrm{i} & -\mathrm{i} \\
 1 & -1 & 1 & -1 & \mathrm{i} & -\mathrm{i} & \mathrm{i} & -\mathrm{i} & 1 & 1 & -1 & -1 & -\mathrm{i} & \mathrm{i} & -\mathrm{i} & \mathrm{i} \\
 1 & -1 & 1 & -1 & \mathrm{i} & -\mathrm{i} & \mathrm{i} & -\mathrm{i} & -1 & -1 & 1 & 1 & \mathrm{i} & -\mathrm{i} & \mathrm{i} & -\mathrm{i} \\
 1 & -1 & 1 & -1 & -\mathrm{i} & \mathrm{i} & -\mathrm{i} & \mathrm{i} & -1 & -1 & 1 & 1 & -\mathrm{i} & \mathrm{i} & -\mathrm{i} & \mathrm{i} \\
 1 & -1 & -1 & 1 & \mathrm{i} & -\mathrm{i} & -\mathrm{i} & \mathrm{i} & \mathrm{i} & -\mathrm{i} & \mathrm{i} & -\mathrm{i} & -1 & -1 & 1 & 1 \\
 1 & -1 & -1 & 1 & -\mathrm{i} & \mathrm{i} & \mathrm{i} & -\mathrm{i} & -\mathrm{i} & \mathrm{i} & -\mathrm{i} & \mathrm{i} & -1 & -1 & 1 & 1 \\
 1 & -1 & -1 & 1 & -\mathrm{i} & \mathrm{i} & \mathrm{i} & -\mathrm{i} & \mathrm{i} & -\mathrm{i} & \mathrm{i} & -\mathrm{i} & 1 & 1 & -1 & -1 \\
 1 & -1 & -1 & 1 & \mathrm{i} & -\mathrm{i} & -\mathrm{i} & \mathrm{i} & -\mathrm{i} & \mathrm{i} & -\mathrm{i} & \mathrm{i} & 1 & 1 & -1 & -1 \\
\end{array}
\right),\\
\tilde{T}_{16}^{(2)}&=\mathrm{diag}\left( 1, 1, 1, 1, 1, 1, -1, -1,1,1,-1,-1, \mathrm{i}, \mathrm{i}, -\mathrm{i},- \mathrm{i}\right),
\end{split}
\end{eqnarray}

\begin{equation}
\tilde{S}_{16}^{(3)}= S_{\mathrm{TC}}\otimes S_{\mathrm{TC}}
,\qquad
\tilde{T}_{16}^{(3)}=T_{\mathrm{TC}}\otimes T_{\mathrm{TC}}.
\end{equation}

\end{subequations}

\end{widetext}

\end{document}